\documentclass[apjl]{emulateapj}
\usepackage{natbib}
\usepackage[backref,breaklinks,colorlinks,citecolor=blue]{hyperref}
\usepackage{color,graphicx}
\usepackage{times,amssymb,amsmath,mathrsfs}

\def\aap{A\&A}                   
\def\apjs{ApJS}                  
\def\apjl{ApJ}                   

\def\MA0{ \mathcal{M}_{{\rm A}, 0} }

\begin{document}

\title[Star forming cores]{Moving mesh simulations of star forming cores in magneto-gravo-turbulence}

\author{Philip Mocz, Blakesley Burkhart, and Lars Hernquist}
\affil{Harvard-Smithsonian Center for Astrophysics, 60 Garden St., Cambridge, MA 02138, USA}
\author{Christopher F. McKee}
\affil{Physics Department and Astronomy Department University of California at Berkeley, Berkeley, CA 94720}
\author{Volker Springel}
\affil{Heidelberger Institut f\"ur Theoretische Studien, Schloss-Wolfsbrunnenweg 35, 69118 Heidelberg, Germany}
\affil{Zentrum f\"ur Astronomie der Universit\"at Heidelberg, Astronomisches
Recheninstitut, M\"onchhofstr. 12-14, 69120 Heidelberg, Germany}
\email{pmocz@cfa.harvard.edu}

\begin{abstract}
Star formation in our Galaxy occurs in molecular clouds that are self-gravitating, highly turbulent, and magnetized. We study the conditions under which cloud cores inherit large-scale magnetic field morphologies and how the field is governed by cloud turbulence. We present four moving-mesh simulations of supersonic, turbulent, isothermal, self-gravitating gas with a range of magnetic mean-field strengths 
characterized by the Alfv\'enic Mach number  $\MA0$, resolving pre-stellar core formation from parsec to a few AU scales. In our simulations with the turbulent kinetic energy density dominating over magnetic pressure ($\MA0>1$), we find that the collapse is approximately isotropic with $B\propto\rho^{2/3}$, core properties are similar regardless of initial mean-field strength, and the field direction on $100$ AU scales is uncorrelated with the mean field. However, in the case of a dominant large-scale magnetic field ($\MA0=0.35$), the collapse is anisotropic with $B\propto\rho^{1/2}$. This transition at $\MA0\sim1$ is not expected to be sharp, but clearly signifies two different paths for magnetic field evolution in star formation.  Based on observations of different star forming regions, we conclude that star formation in the interstellar medium may occur in both regimes.
Magnetic field correlation with the mean-field extends to smaller scales as $\MA0$ decreases, making future ALMA observations useful for constraining $\MA0$ of the interstellar medium.
\end{abstract}


\section{Introduction}

Magnetic fields and turbulence are known to play key roles in star formation \citep{1981MNRAS.194..809L,1987ARA&A..25...23S,1993prpl.conf..327M,2010ApJ...725..466C,2012ARA&A..50...29C}, and
both compete against the self-gravity of the gas and strongly affect the dynamics of collapse. The relative importance of one over the other may have significant consequences for how pre-stellar cores collapse, including whether the collapse: (1) is isotropic, (2) is self-similar, and (3) has magnetic field lines that shape filamentary structure or turbulence that shape the field lines.

Turbulent motions in the interstellar medium are known to be highly supersonic \citep{1981MNRAS.194..809L, burkhart2015}, with giant molecular clouds (GMCs) having sonic Mach numbers $\mathcal{M}_{\rm s}\sim 10$, meaning that turbulent pressure greatly exceeds thermal pressure. The magnetic field is also known to be important.  Often a coherent mean-field can be measured on large (parsec) scales and density structures in the ISM are filamentary, and diffuse clouds are thought to be assembled by flows along magnetic field lines \citep{2010ApJ...725..466C}.
The relative importance between the turbulent kinetic energy density $E_{{\rm k},{\rm turb}}$ and the magnetic pressure $P_B$ can be characterized by the Alfv\'enic Mach number $\mathcal{M}_{\rm A} = (E_{{\rm k},{\rm turb}} / P_B)^{1/2}$, where super-Alfv\'enic ($\mathcal{M}_{\rm A}>1$) signifies turbulence dominance and sub-Alfv\'enic ($\mathcal{M}_{\rm A}<1$) signifies magnetic pressure dominance.

How gravitational collapse occurs in molecular clouds in our Galaxy is debated both observationally and theoretically.
Observationally, Zeeman measurements of interstellar magnetic 
field strengths give the mass-averaged line-of-sight field, $\langle B_z\rangle_M$, in the telescope beam. 
\cite{2010ApJ...725..466C} collected Zeeman observations of both atomic and molecular gas and used Bayesian analysis to infer that the mass-weighted total field
satisfied $\langle B\rangle_M\propto \langle \rho\rangle^{0.65}$, where $\langle \rho\rangle$ is the mean density of the gas in the observed region, in gas denser than $n_{\rm H} =300\,{\rm cm}^{-3}$, almost all of which is molecular. 
\cite{2015MNRAS.452.2500L} confirmed this result by showing that the \cite{2010ApJ...725..466C} data on molecular gas implies
$\langle B\rangle_M\propto \langle \rho\rangle^\alpha$ with $\alpha=0.64 \pm 0.13$.
A non-turbulent pre-stellar core undergoing homologous contraction with a frozen-in field has $B\propto \rho^{2/3}$ (\citealt{1965QJRAS...6..161M};  the core must be pre-stellar since non-ideal MHD effects become important in the vicinity of a protostar).
The observations are also consistent with the model of fast turbulent reconnection diffusion of the magnetic field relative to the free fall time of collapse \citep{Lazarian2012}.
In their ideal MHD simulations of a turbulent, self-gravitating gas, 
\cite{2015MNRAS.452.2500L} compute the scaling relation taking density-averaged magnetic fields, as observers would from Zeeman measurements, and apply observational effects such as convolution with the beam, for the 100 most massive
clumps in their simulation: they found a result for $\langle B\rangle_M$ consistent with the observations of \cite{2010ApJ...725..466C}. They considered Alfv\'enic ($\MA0=1$) and super-Alfv\'enic ($\MA0=10$) turbulence only, where $\MA0$ is based on the mean magnetic field.

Alternatively, \cite{2015MNRAS.451.4384T} suggest that the \cite{2010ApJ...725..466C} observations better support a $B\propto\rho^{1/2}$ scaling based on re-estimating observational uncertainties in the determination of cloud and core densities, signaling anisotropic collapse with dynamically important magnetic fields. Previous theoretical work  \citep{mestel1966,1976ApJ...207..141M,1976ApJ...206..753M} on the problem of the equilibrium of self-gravitating, isothermal, strongly magnetized clouds in the ISM finds that the ratio of magnetic and gas pressure should remain close to unity near the center of the cloud, $B_{\rm c}\propto\rho^{1/2}$.
Axisymmetric calculations of gravitational collapse in a strong magnetic field under the influence of ambipolar diffusion give $B_{\rm c}\propto\rho_{\rm c}^\kappa$ with $\kappa < 1/2$ \citep{1994ApJ...425..142C}.
It should be borne in mind, however, that $\kappa$ in the relation between $B_{\rm c}$ and $\rho_{\rm c}$ may be independent of the power-law index in the observed $\langle B\rangle_M(\langle \rho\rangle)$ relation. 
Ideal MHD simulations of star forming cores in turbulent environments with weak mean-field ($\MA0\gg 1$) by \cite{2011ApJ...731...59C,2012ApJ...750...13C} find a scaling close to $B_{\rm c}\propto\rho_{\rm c}^{0.4-0.5}$ when examining the distribution magnetic fields and densities of the gas cells in their simulation.
The scaling relation, simply deduced from a cell-by-cell determination, is different to the $B\propto\rho^{2/3}$ scaling of \cite{2015MNRAS.452.2500L} using density-averaged line-of-sight fields.
Recent observations of the massive star forming region NGC 6334 \citep{2015Natur.520..518L} on $100$--$0.01$~pc scales shows self-similar hourglass-like magnetic field structure with $B\propto\rho^{0.41}$. We note, however, that the magnetic field values in \cite{2015Natur.520..518L} are inferred using various approximate methods, as opposed to being directly measured by Zeeman splitting as in \cite{2010ApJ...725..466C}, so the errors associated with the slope of the relation are much larger. Hourglass morphologies, suggesting dynamically important magnetic fields, have been observed in a number of other cores as well \citep{2006Sci...313..812G,2013ApJ...769L..15S,2009ApJ...707..921R,2009ApJ...700..251T,2014ApJ...794L..18Q}; however, such morphologies also occur in the simulations of \cite{2015MNRAS.452.2500L}, which have $\MA0=1$.

These observations and theoretical predictions have important implications for the influence and transport of the magnetic field during collapse.  However, only in recent years have numerical simulations advanced to the point where
theoretical predictions can be tested and observational parameters/physics reproduced.  This is, in part, due to the extreme numerical expense of simulating many orders of magnitude in spatial scale and density range to study the collapse of a parsec scale cloud down to the AU scale disk where a star is born.

In this paper, we present novel moving mesh \textsc{Arepo} simulations of the collapse of pre-stellar cores in supersonic, turbulent, isothermal, magnetized environments, exploring the effect of the mean magnetic field-strength (an invariant of ideal MHD). These numerical simulations self-consistently resolve star formation in a large-scale turbulent environment (5pc) down to a few AU scale and are well-situated to help improve our theoretical understanding of the star formation process, interpret observations, and constrain the regimes in which star formation occurs. The moving mesh numerical framework allows us to efficiently explore a range of initial field strengths and resolve the core collapse down to AU scales, relevant for upcoming  sub-arcsecond spatial resolution observations by the Atacama Large Millimeter Array (ALMA; \citealt{2015ApJ...808L...3A}). We are able to resolve the cores by a factor of $>8$ improvement in spatial resolution compared to similar studies that use adaptive mesh refinement (AMR) \citep{2011ApJ...731...59C,2012ApJ...750...13C,2015MNRAS.452.2500L}, until  the approximate isothermal condition breaks down and the core is expected to continue adiabatic collapse on smaller scales. These simulations are particularly relevant to understanding the morphologies of Class 0 protostars.

The paper is organized as follows. In \S~\ref{sec:sim} we describe our numerical methods and simulation setup.  We present projected morphologies in \S~\ref{sec:proj}. We investigate the density dependence of the magnetic field in the pre-stellar cores in \S~\ref{sec:Brho} and compute their radially averaged profiles in \S~\ref{sec:prof}. We discuss our findings in \S~\ref{sec:disc} and our main conclusions in \S~\ref{sec:conc}.

\begin{table*}
\caption{Simulation initial conditions and final configurations.}
\begin{center}
\begin{tabular}{ccccccccccc}
\hline
\hline
sim. & $\beta_0$ & $\MA0$ & $\mathcal{M}_{{\rm s}}$ & $\mu_{\Phi,0}$ & comment & 
\begin{tabular}{@{}c@{}}
first core \\ $t_{\rm collapse}$ ($t_{\rm ff}$)
\end{tabular}
  & $\alpha$ ($B\propto\rho^{\alpha}$)
&  $\mu_{\Phi}$ & 
\begin{tabular}{@{}c@{}}
mean-field \\ aligned within $30^\circ$ \\  $10^6~{\rm AU}$ to $(\cdot)~{\rm AU}$
\end{tabular} & 
\begin{tabular}{@{}c@{}}
fraction of gas \\ at $10^4~{\rm AU}$ scale \\   with field aligned \\ within $30^\circ$
\end{tabular}
\\
\hline
1 & 25     & 35   & 10 & 80  & very weak field & 0.12 & $0.64\pm0.02$ & 12.7 & $10^{2.0}$ & 0.00\\
2 & 0.25   & 3.5  & 10 & 8   & weak field      & 0.16 & $0.66\pm.02$  & 16.5 & $10^{3.7}$ & 0.23\\
3 & 0.028  & 1.2  & 10 & 2.7 & moderate field  & 0.17 & $0.67\pm.01$  & 12.1 & $10^{4.0}$ & 0.35\\
4 & 0.0025 & 0.35 & 10 & 0.8 & strong field    & 0.37 & $0.55\pm.02$  & 5.8 & $10^{5.5}$ & 0.86\\
\hline
\hline
\end{tabular}
\end{center}
\label{tbl:sims}
\end{table*}

\begin{figure*}
\begin{center}
\begin{tabular}{ccc}
{\Large $\MA0 = 35$} &
{\Large $\MA0 = 3.5$} & \\
\includegraphics[width=0.4\textwidth]{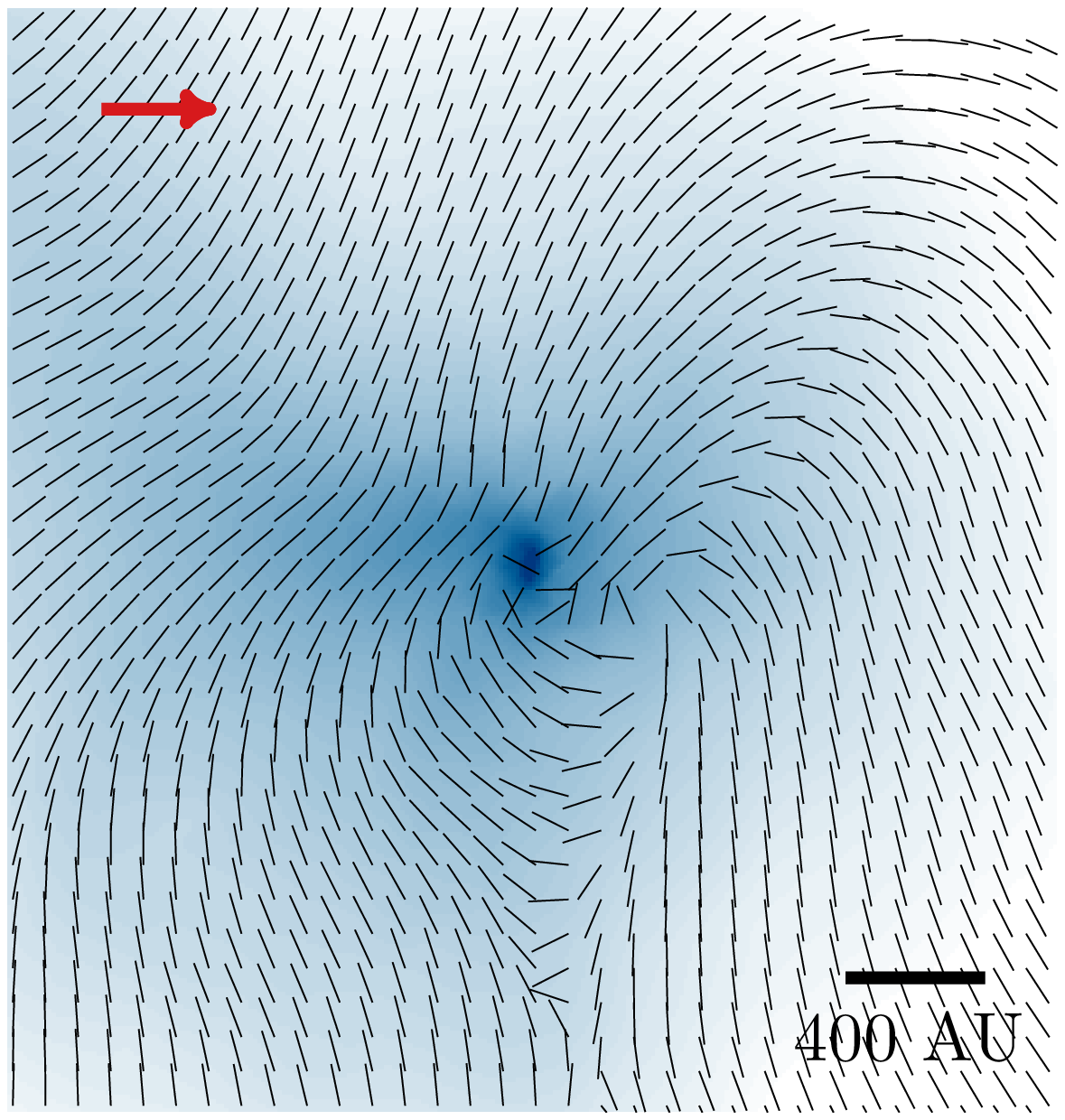} &
\includegraphics[width=0.4\textwidth]{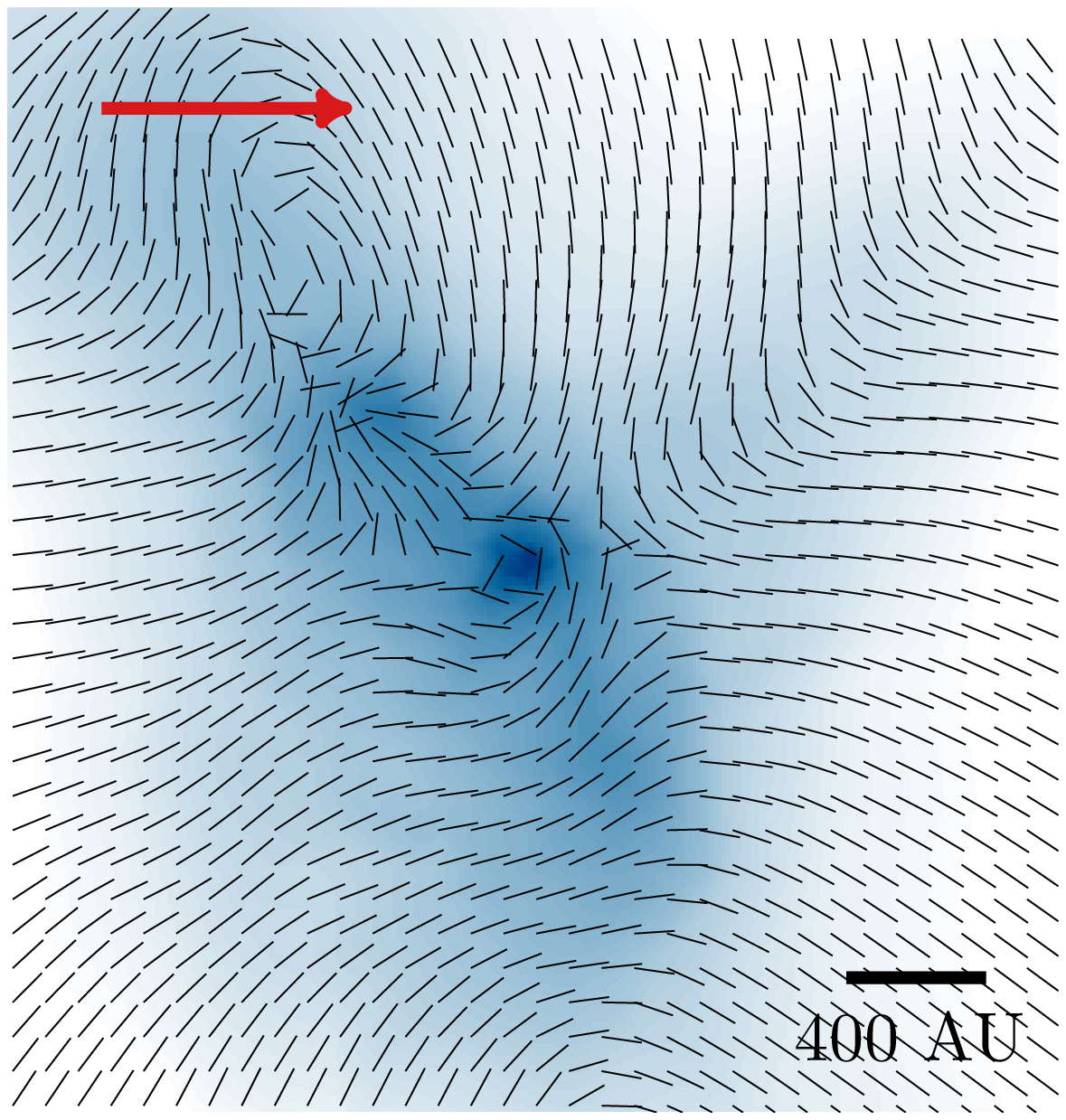}  &
\includegraphics[width=0.105\textwidth]{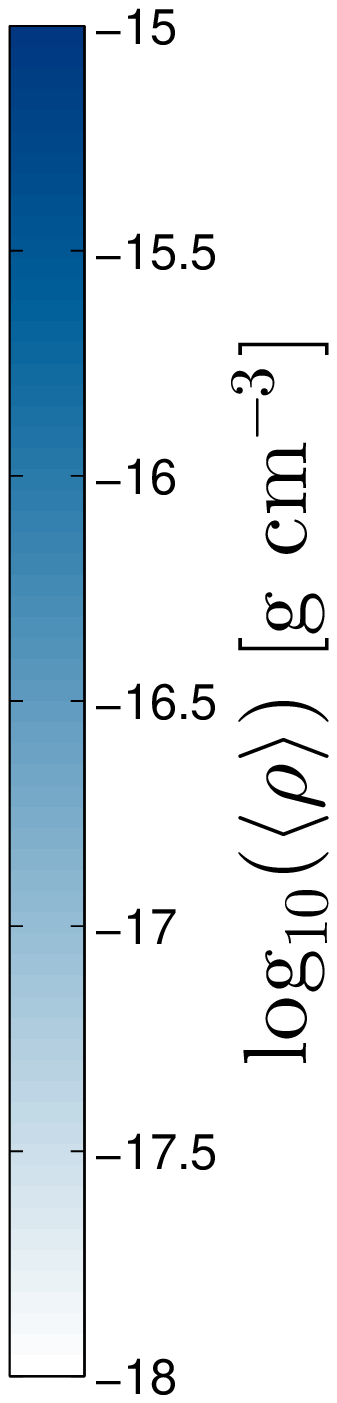}\\
& & \\
{\Large $\MA0 = 1.2$} & 
{\Large $\MA0 = 0.35$} & \\
\includegraphics[width=0.4\textwidth]{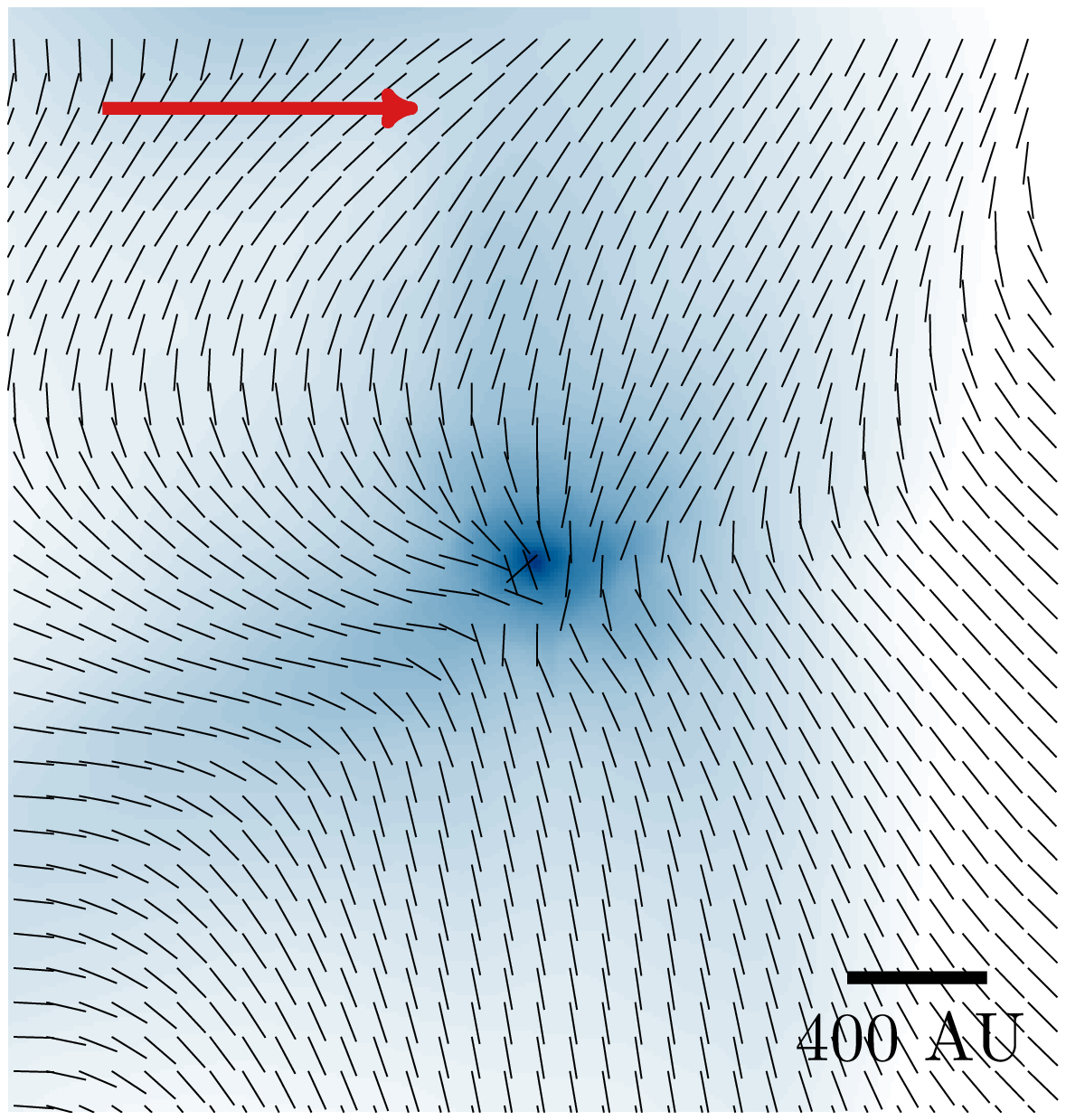} &
\includegraphics[width=0.4\textwidth]{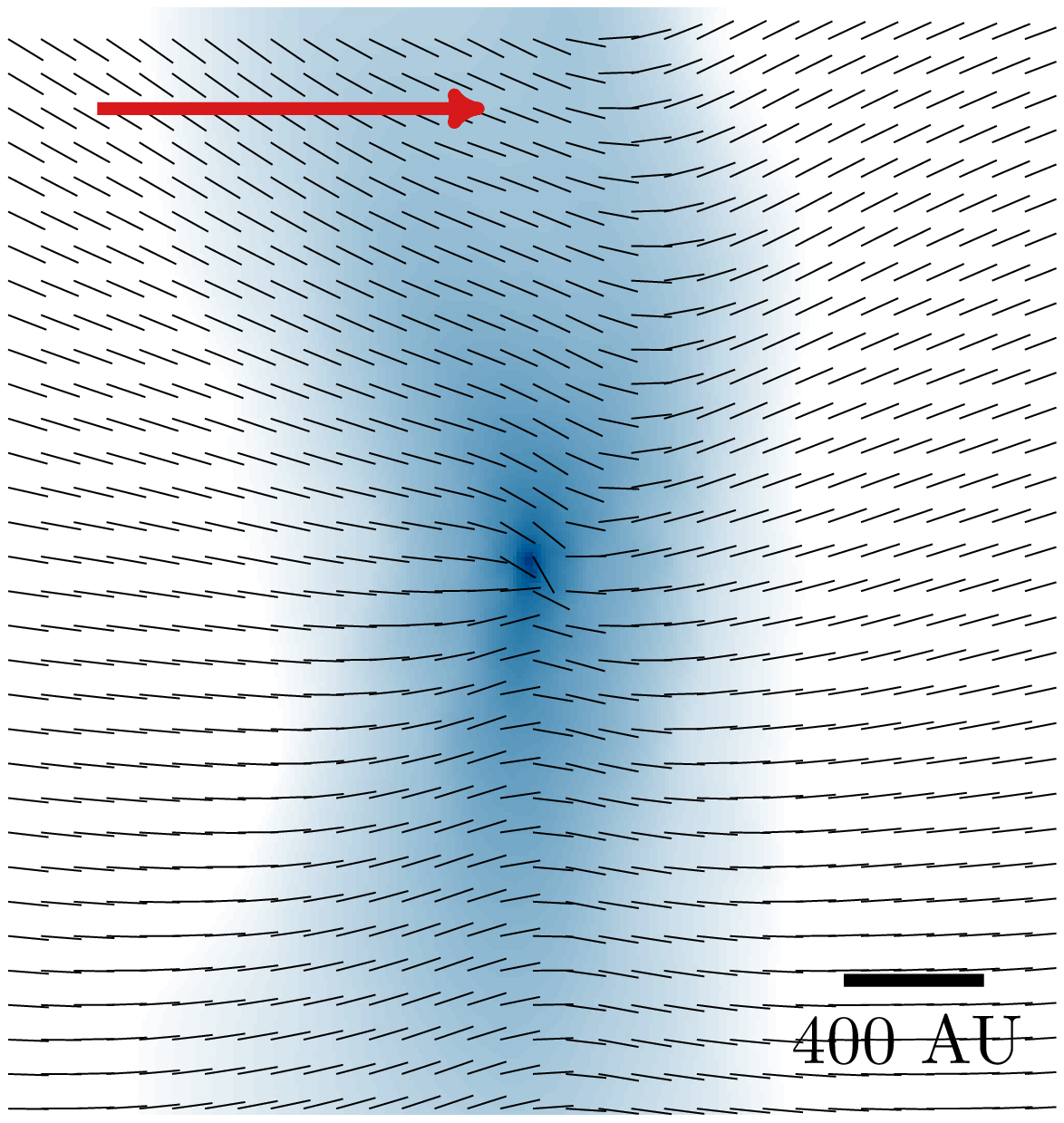} &
\includegraphics[width=0.105\textwidth]{colorbar.eps}
\end{tabular}
\end{center}
\caption{Density-averaged line-of-sight magnetic field and projected gas densities zoomed in on a core in each of the simulations. The arrow in the top left represents the initial magnetic field.}
\label{fig:sim_proj}
\end{figure*}

\begin{figure*}
\begin{center}
\begin{tabular}{ccc}
{\Large $\MA0 = 35$} &
{\Large $\MA0 = 3.5$} & \\
\includegraphics[width=0.4\textwidth]{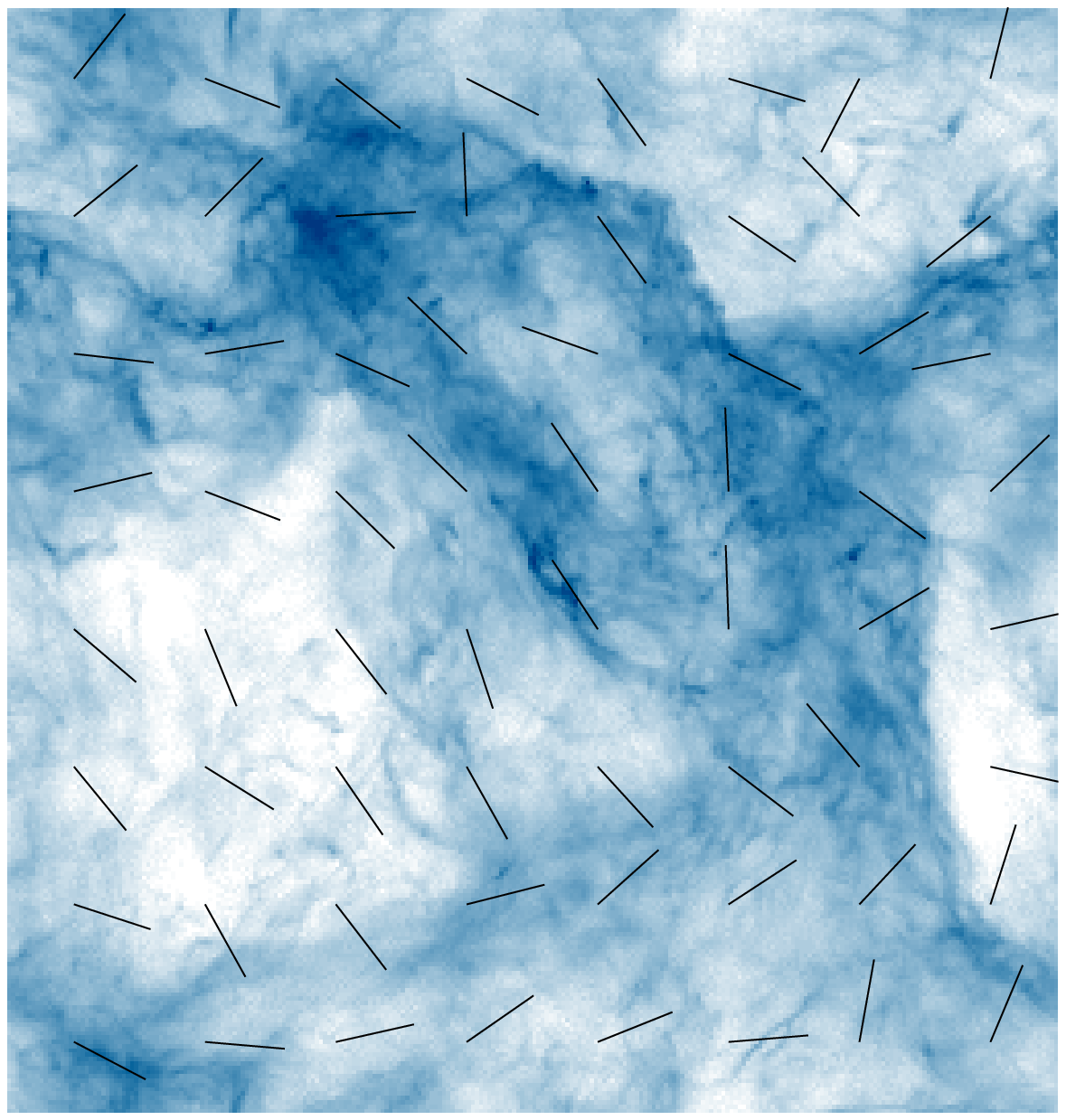} &
\includegraphics[width=0.4\textwidth]{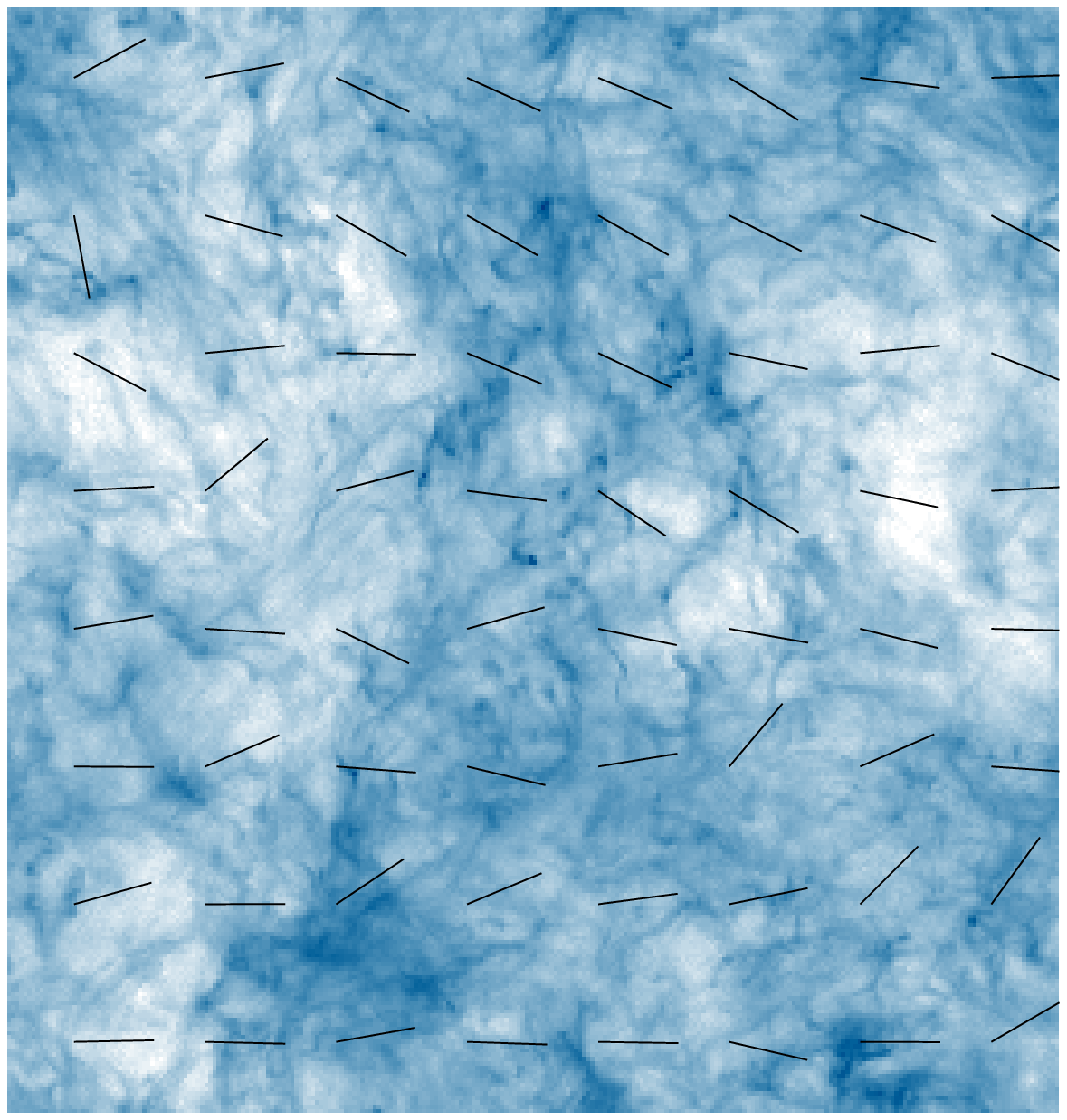}  &
\includegraphics[width=0.105\textwidth]{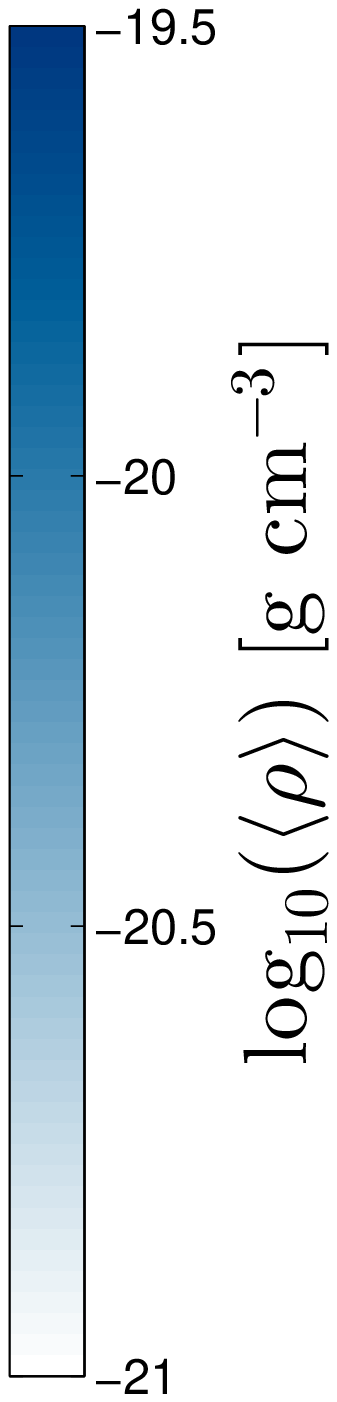}\\
& & \\
{\Large $\MA0 = 1.2$} & 
{\Large $\MA0 = 0.35$} & \\
\includegraphics[width=0.4\textwidth]{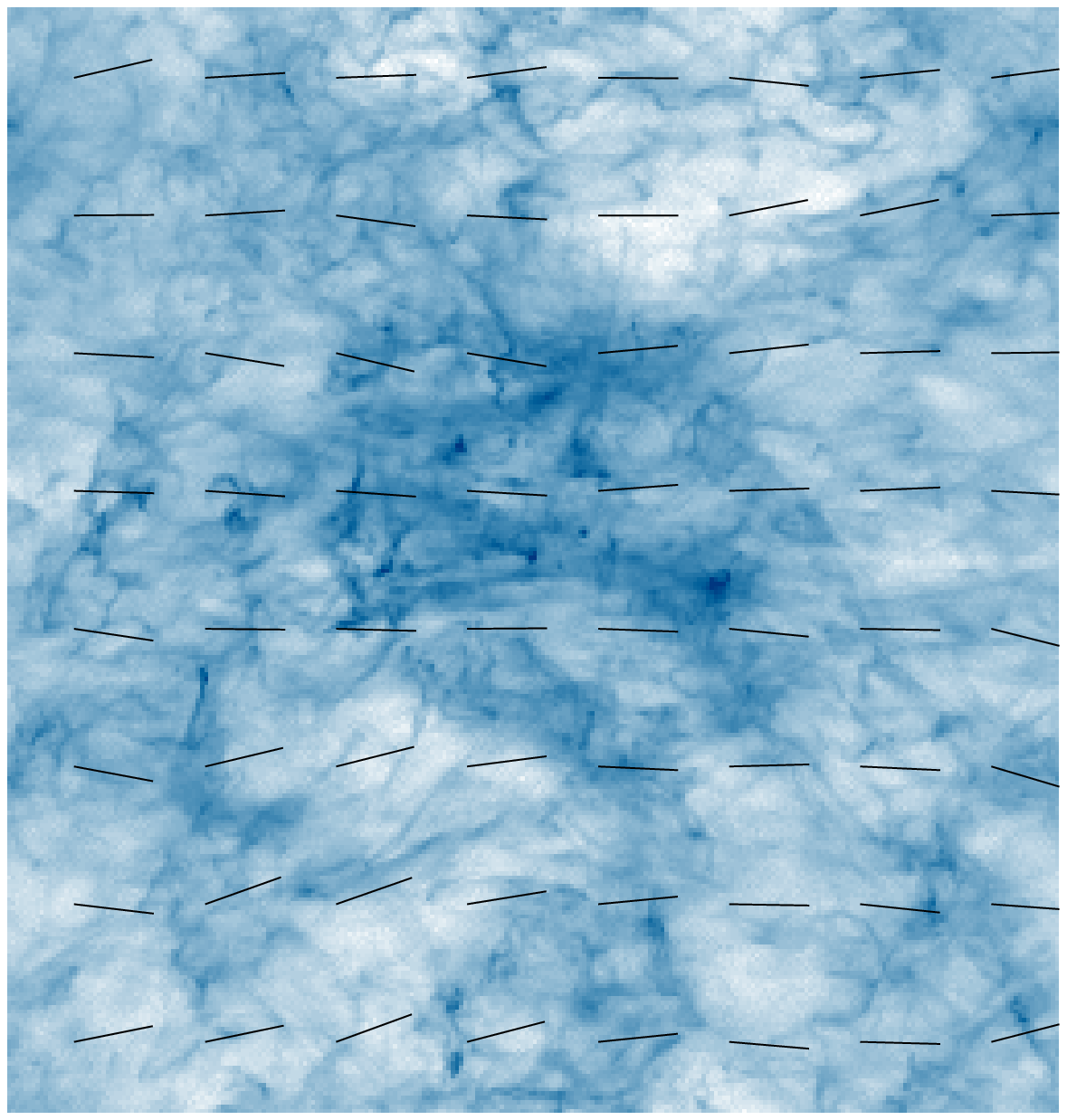} &
\includegraphics[width=0.4\textwidth]{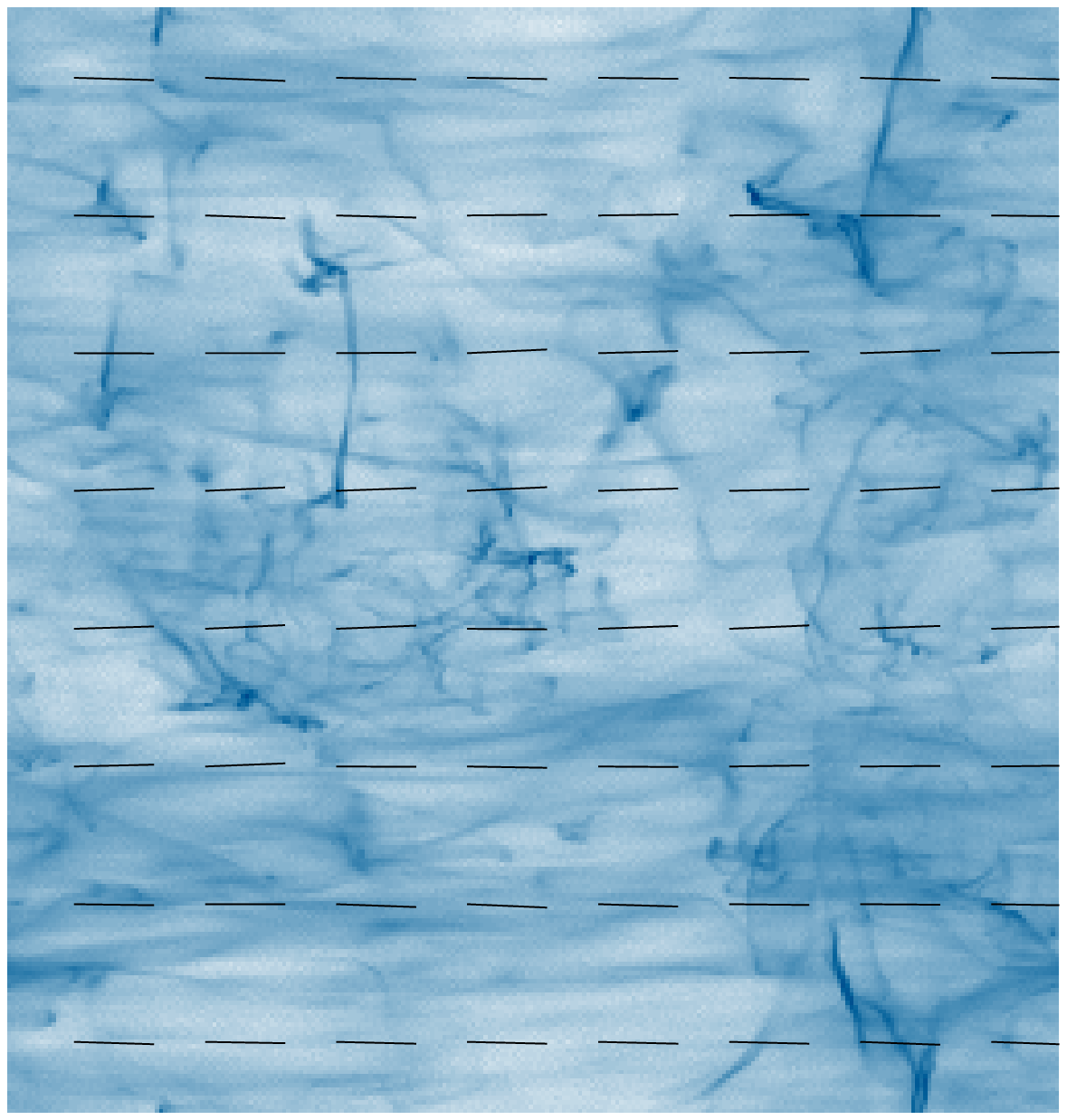} &
\includegraphics[width=0.105\textwidth]{boxcb.eps}
\end{tabular}
\end{center}
\caption{Density-averaged line-of-sight magnetic field and projected gas densities of the entire simulation domain ($5.2~{\rm pc}$) for the 4 simulations. Each box is centered on the corresponding pre-stellar cores shown in Fig.~\ref{fig:sim_proj}.}
\label{fig:sim_box}
\end{figure*}

\begin{figure}
\begin{center}
\includegraphics[width=0.47\textwidth]{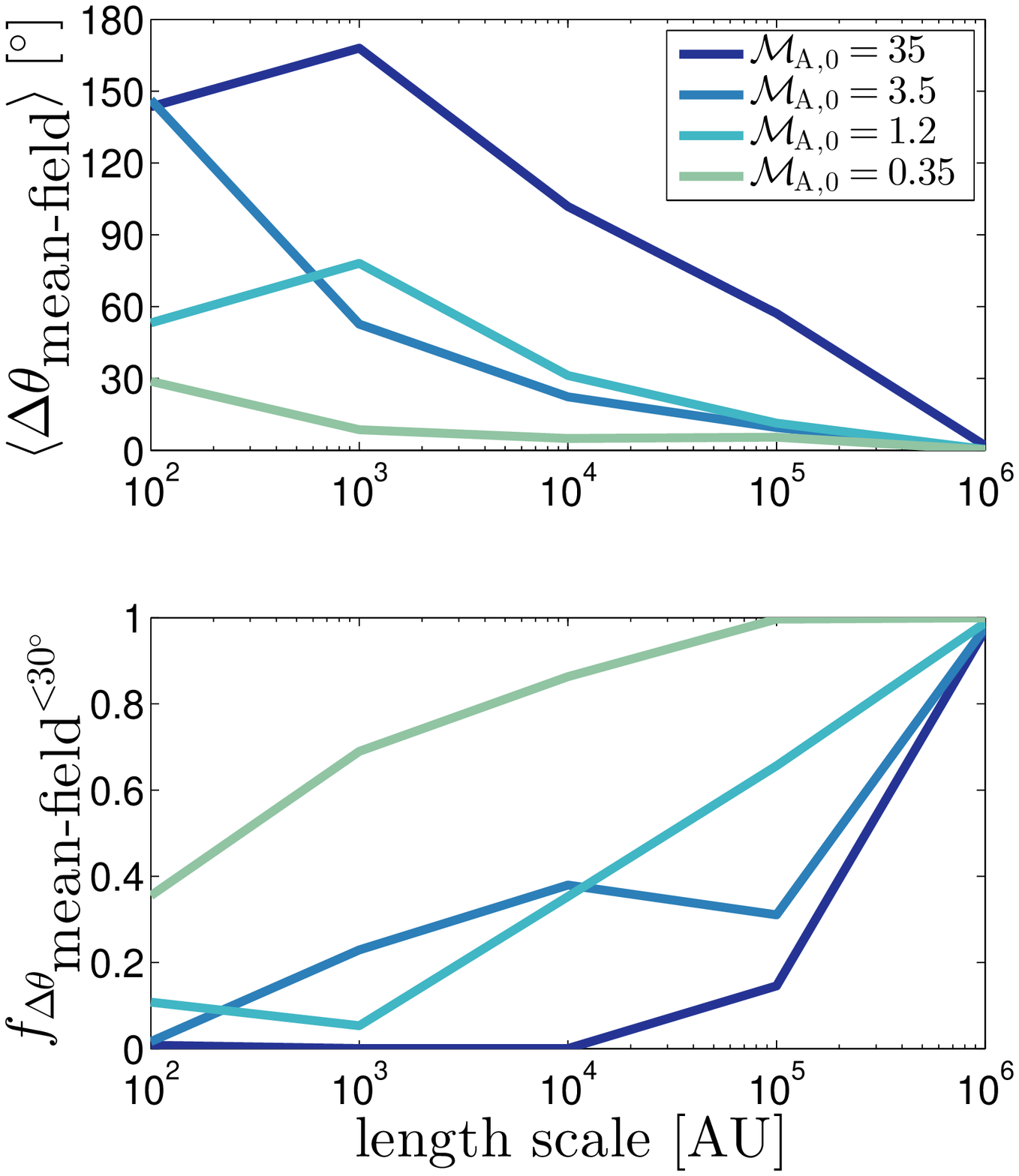}
\end{center}
\caption{(\textit{Top}) Deviation of the mean-magnetic field averaged over some length-scale around the core from the large (pc) scale field. (\textit{Bottom}) Fraction of gas inside a sphere of radius the length scale with the magnetic field aligned within $30$~degrees of the large-scale mean-field value.}
\label{fig:meanField}
\end{figure}

\section{Simulations}
\label{sec:sim}

We simulate the collapse of star forming cores under self-gravity in a turbulently-driven interstellar medium environment. Our simulations are performed using the moving-mesh quasi-Lagrangian \textsc{Arepo} code \citep{2010MNRAS.401..791S}. The moving mesh automatically adapts to the geometry of the physical system, and keeps the mass-resolution of each cell approximately constant. The code solves the ideal MHD equations, for which we have recently implemented \citep{2016MNRAS.463..477M} an unstructured vector potential constrained transport \citep{1966ITAP...14..302Y,1988ApJ...332..659E} solver to maintain the divergence-free property of the magnetic field. We accurately capture shocks via an HLLD \citep{2005JCoPh.208..315M} Riemann solver. Self-gravity is calculated using a Tree-Particle-Mesh scheme \citep{1995ApJS...98..355X}. Turbulence is driven solenoidally in Fourier space on the largest spatial scales using an Ornstein-Uhlenbeck process \citep{2010A&A...512A..81F,2012MNRAS.423.2558B,2015MNRAS.450.4035F}.

We run four isothermal simulations, representing part of a GMC, with different initial mean-field strengths, $B_0$. Turbulence is characterized by the sonic Mach number $\mathcal{M}_{\rm s} = v_{\rm rms}/c_{\rm s} = 10$. The cloud has Virial parameter: $\alpha_{\rm vir} = 5 v_{\rm rms}^2 (L/2) /(3G M_0)=1/2$.
 The physical parameters of the simulations (assuming a mass per hydrogen of $1.4~{\rm amu}$) can be scaled as:
\begin{equation}
\begin{array}{l}
L_0 = 5.2 \left(\frac{c_s}{0.2~{\rm km}~{\rm s}^{-1}}\right)
\left(\frac{n_H}{1000~{cm}^{-3}}\right)^{-1/2}
\left(\frac{\mathcal{M}_{\rm s}}{10}\right)
 ~{\rm pc} \\
B_0 = 1.2,12,36,120 \left(\frac{c_s}{0.2~{\rm km}~{\rm s}^{-1}}\right)
\left(\frac{n_H}{1000~{cm}^{-3}}\right)^{1/2}
 ~{\rm \mu G} \\
M = 4860 \left(\frac{c_s}{0.2~{\rm km}~{\rm s}^{-1}}\right)^3
\left(\frac{n_H}{1000~{\rm cm}^{-3}}\right)^{-1/2} 
\left(\frac{\mathcal{M}_{\rm s}}{10}\right)^3 
 M_\odot \\
 \end{array}
\end{equation}
where $L_0$ is the size of the periodic box with total mass of $M$. We scale the simulations to physical units using sound speed $c_{\rm s}=0.2~{\rm km}~{\rm s}^{-1}$ and hydrogen density $n_{\rm H} = 1000~{\rm cm}^{-3}$.\footnote{
We note our choice of physical scaling makes the cloud 
fall on the observed line width-size scaling relation for molecular clouds in our Galaxy:
$\sigma_{\rm nt} = \sigma_{\rm pc} R_{\rm pc}^{1/2}$, with $\sigma_{\rm pc}\simeq 0.72~{\rm km}~{\rm s}^{-1}$ \citep{2007ARA&A..45..565M}. Here $R_{\rm pc}=(L_0/2)/(1~{\rm pc})$ and $\sigma_{\rm nt} = \mathcal{M}_{\rm s}c_{\rm s}/\sqrt{3}$ (as in \citealt{2010ApJ...720.1612M}).
}

The simulation initial conditions, ranging from very weak seed fields to strong fields that surpass the turbulent kinetic energy density, are summarized in Table~\ref{tbl:sims}. The plasma beta parameter ($\beta = P_{\rm gas}/P_B = 8\pi P_{\rm gas}/B^2$) describes whether gas pressure ($\beta>1$) or magnetic pressure ($\beta<1$) dominates. Similarly, the Alfv\'enic Mach number $\mathcal{M}_{\rm A}$ describes whether the turbulent kinetic energy density ($\mathcal{M}_{\rm A}>1$) or magnetic pressure ($\mathcal{M}_{\rm A}<1$) dominates. Note $\beta = 2\mathcal{M}_{\rm A}^2/\mathcal{M}_{\rm s}^2$. Each simulation uses $256^3$ cells, corresponding to a mass resolution of $8\cdot 10^{-5}~{M_\odot}$.

In the simulations, we first drive turbulence until quasi-steady state is established after a couple of eddy-turnover times. Then, we turn on self-gravity (continuing to drive turbulence) which eventuates in the collapse of cores on the order of the free-fall time. We follow the collapse for a fraction of the free-fall time, until we form cores that are resolved at the level of a few AU at their centers. We stop the simulations at this point because the time-stepping becomes prohibitively expensive and we lack mass-resolution to further resolve the collapse. Furthermore the isothermal assumption is expected to break down beyond these scales, because opacity increases and is able to trap heat and hence the collapse continues adiabatically. Our simulation strategy is to accurately resolve the first core that forms to small (AU) scales. The core centers are identified by determining local minima in the gravitational potentials.
Table~\ref{tbl:sims} lists the times at which these pre-stellar cores form which we analyze.
A discussion on the other (less-collapsed) cores that result from turbulent fragmentation are discussed in  Appendix~\ref{sec:allcores}.

The mass-to-flux ratio is parametrized in dimensionless form as $\mu_{\Phi,0}\equiv M_0/M_\Phi$ ($M_\Phi$ is the magnetic critical mass that can undergo gravitational collapse) assesses the relative strength of the magnetic field and gravity. For the box in its initial uniform, static state, it can be defined as $\mu_{\Phi,0}\sim \sqrt{5\pi/3}\MA0$ (Equation 27 of \citealt{2010ApJ...720.1612M}). Thus our four simulation initial conditions are characterized by $\mu_{\Phi,0}=80, 8, 2.7, 0.8$. The strong-field box is estimated to be slightly sub-critical, but we find collapse if turbulence is switched on (the core in this case forms later, at $\sim0.4t_{\rm ff}$, as opposed to $\sim 0.2t_{\rm ff}$ in the weaker-field simulations). Being able to form a core in the sub-critical case is consistent with the picture of reconnection diffusion, which acts to reduce the mass-to-flux ratio in the core \citep{Lazarian2012}.

Our simulations are similar to the setup of 
\cite{2011ApJ...731...59C,2012ApJ...750...13C} and \cite{2015MNRAS.452.2500L}. However, \cite{2011ApJ...731...59C,2012ApJ...750...13C} assume a very weak magnetic field. \cite{2015MNRAS.452.2500L} has one weak-field and one moderate ($\MA0=1$) setup. Furthermore, these previous AMR simulations do not resolve the collapse down to a few AU scales: the effective resolutions of \cite{2012ApJ...750...13C} and \cite{2015MNRAS.452.2500L} are $120$~AU and $500$~AU respectively. Our smallest cells have an effective diameter of $d = 2\left(3V/(4\pi)\right)^{1/3}\simeq 4~{\rm AU}$ (where $V$ is the volume of the cell), corresponding to adding $\geq 3$ extra refinement levels to these AMR simulations.

\textsc{Arepo} uses a refinement/derefinement strategy to ensure all cells maintain their initial mass to within a factor of 2. Additionally, further refinement is used to ensure the Truelove criterion is met to accurately resolve the local Jeans length $\lambda_{\rm J}=\left(c_{\rm s}^2 \pi / (G\rho)\right)^{1/2}$ \citep{1997ApJ...489L.179T} and prevent artificial fragmentation, whose implementation is described in \cite{2015MNRAS.446.2380B}.

\begin{figure*}
\begin{center}
\begin{tabular}{cc}
{\Large $\MA0 = 35$} &
{\Large $\MA0 = 3.5$} \\
\includegraphics[width=0.45\textwidth]{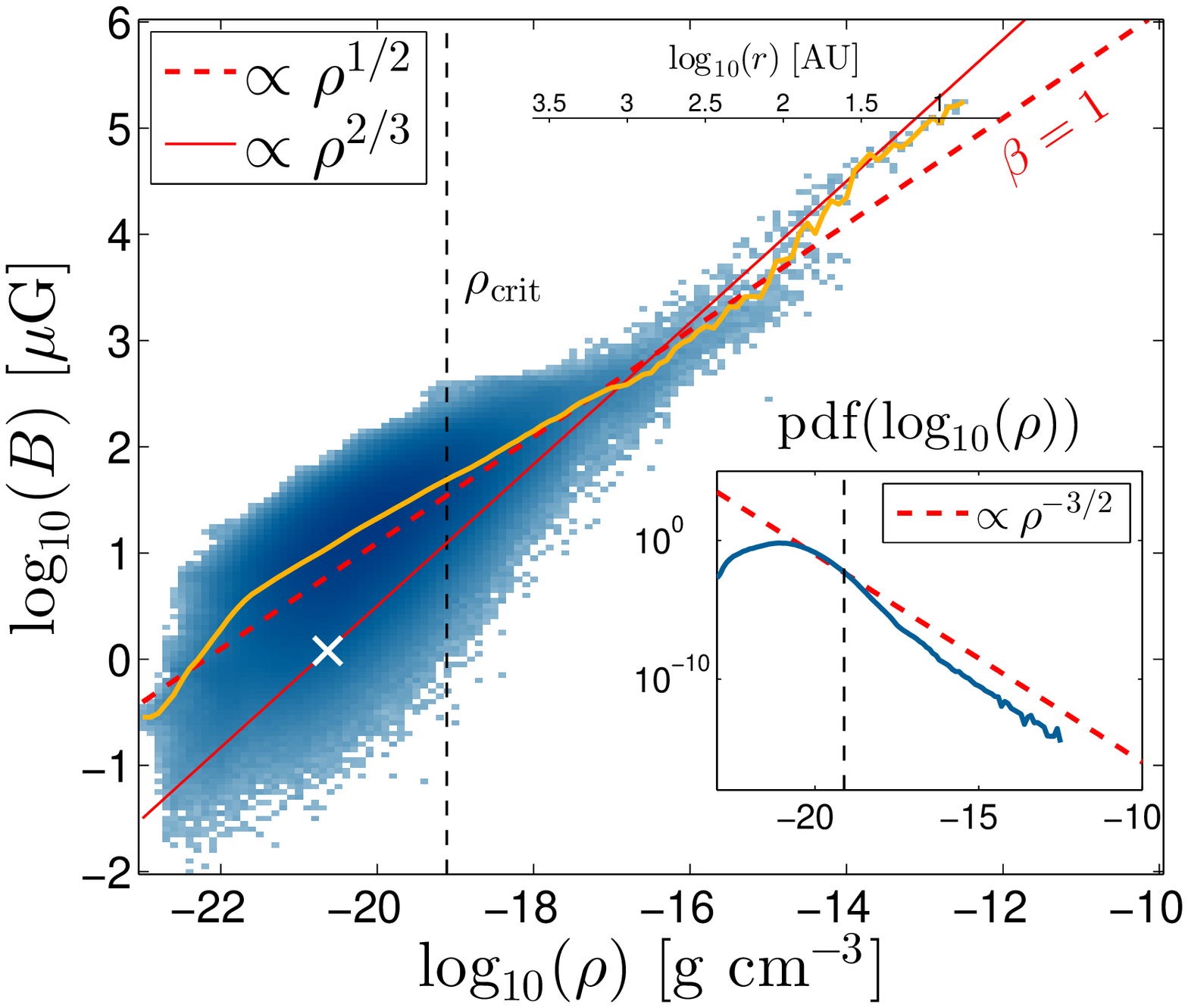} &
\includegraphics[width=0.45\textwidth]{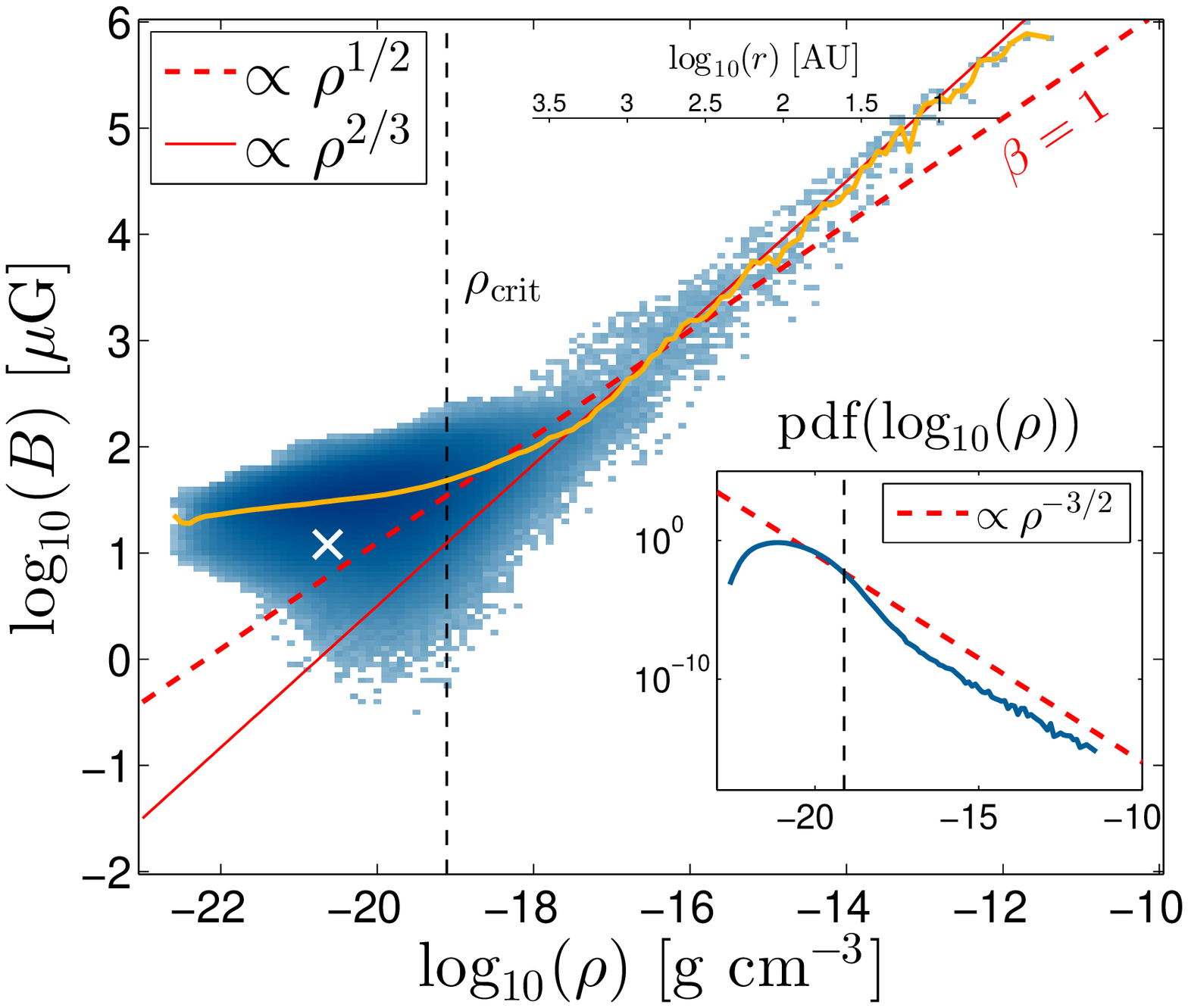} \\
& \\
{\Large $\MA0 = 1.2$} & 
{\Large $\MA0 = 0.35$} \\
\includegraphics[width=0.45\textwidth]{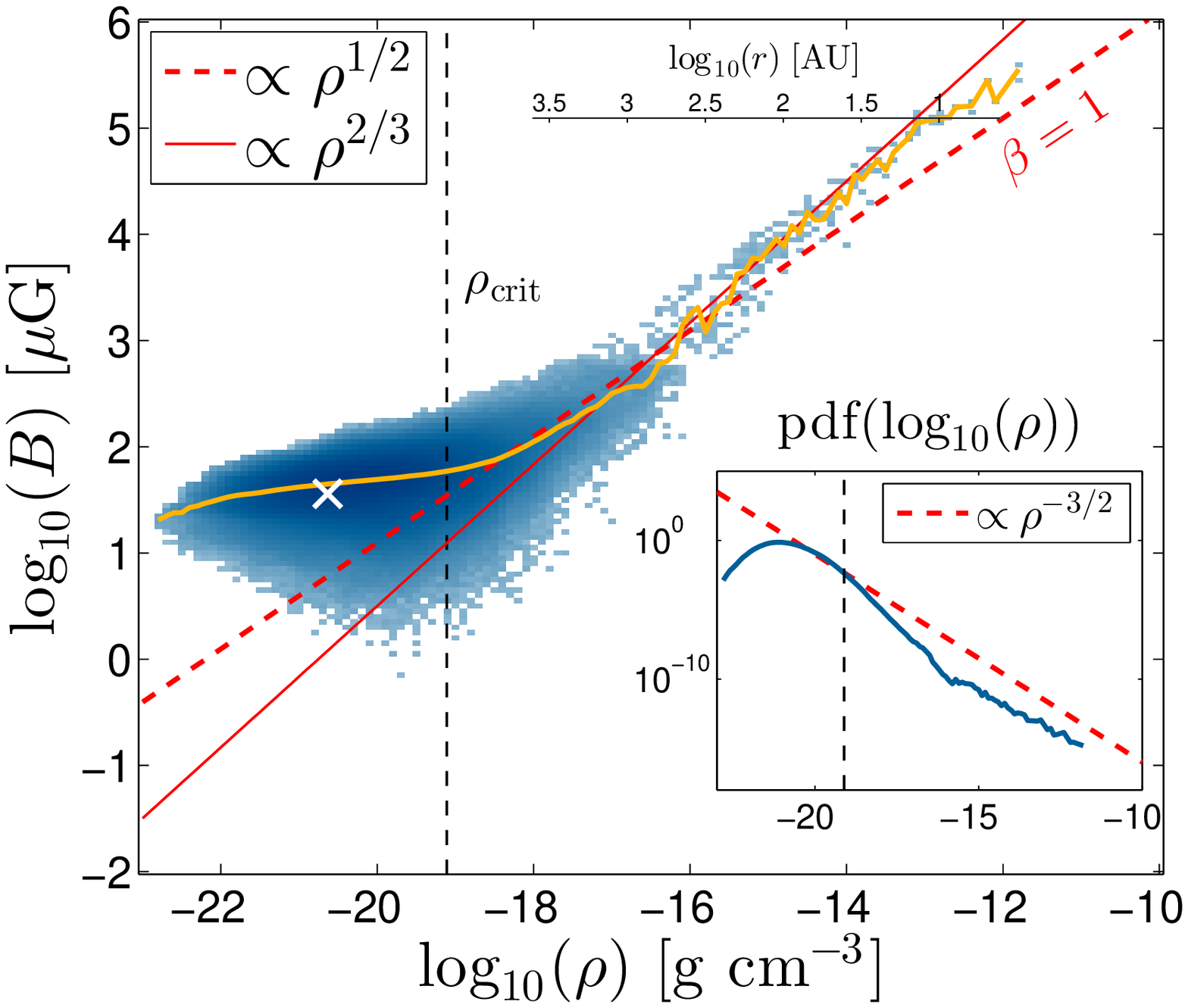} &
\includegraphics[width=0.45\textwidth]{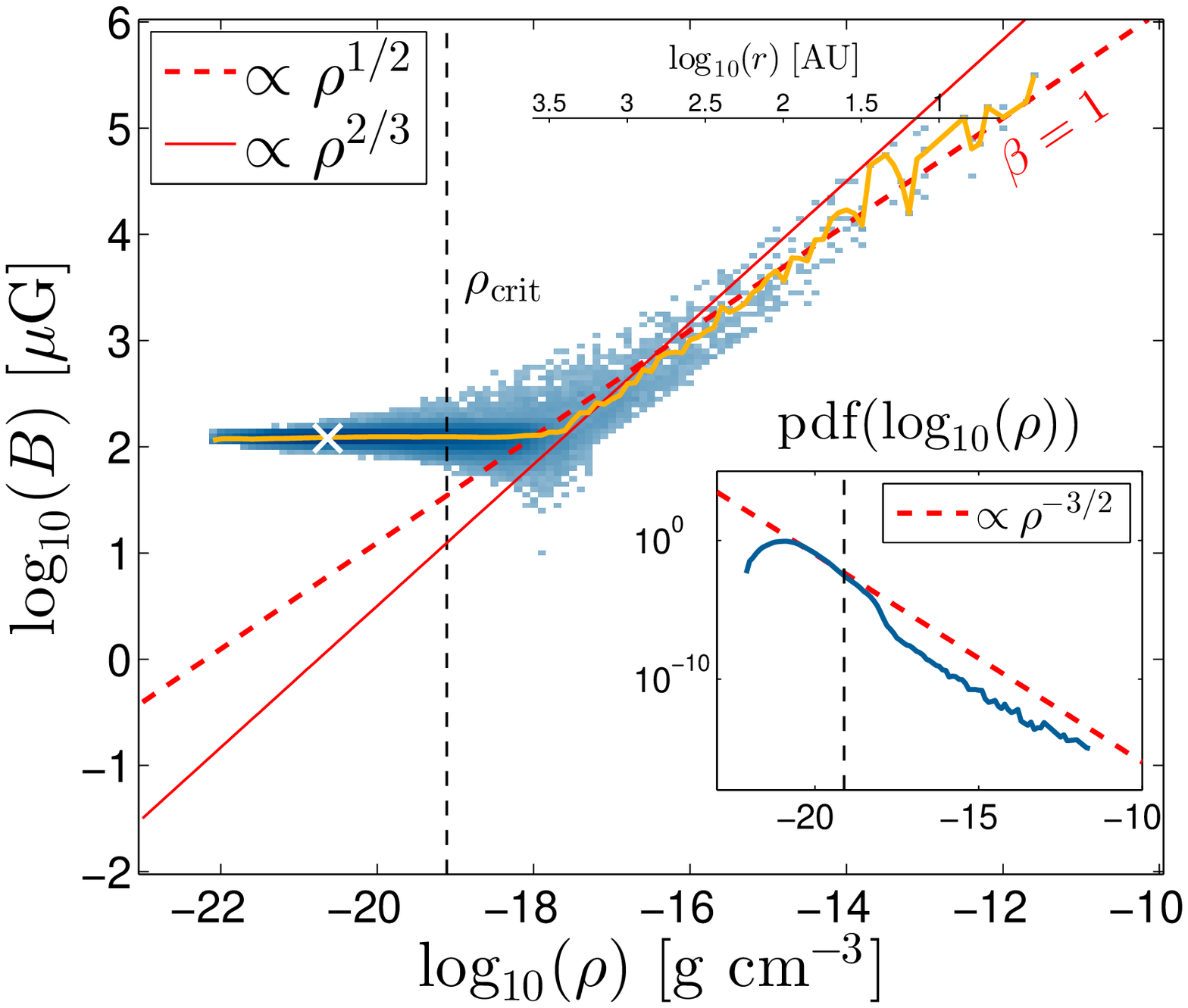}
\end{tabular}
\end{center}
\caption{Magnetic field/density phase diagram for the four simulations. The yellow line shows the average field for each logarithmic density bin. The color-scale shows the mass of gas in each bin of the phase diagram. The red dashed line shows a $\rho^{1/2}$ scaling (with normalization such that $\beta=1$) while the red solid line shows a $\rho^{2/3}$ scaling of isotropic collapse. The white `x' marks the value of the initial magnetic field strength and density in the box. Also marked is the critical density threshold $\rho_{\rm crit}$ where the gas pressure equals the background turbulent pressure. Finally, an approximate conversion from density to a length scale is shown in the collapsed region, based on a $\rho(r) = c_{\rm s}^2G^{-1}r^{-2}$ profile. The inset shows the PDF of the collapsed gas density.}
\label{fig:sim_Brho}
\end{figure*}

\begin{figure*}
\begin{center}
\begin{tabular}{cc}
{\Large $\MA0 = 35$} &
{\Large $\MA0 = 3.5$} \\
\includegraphics[width=0.45\textwidth]{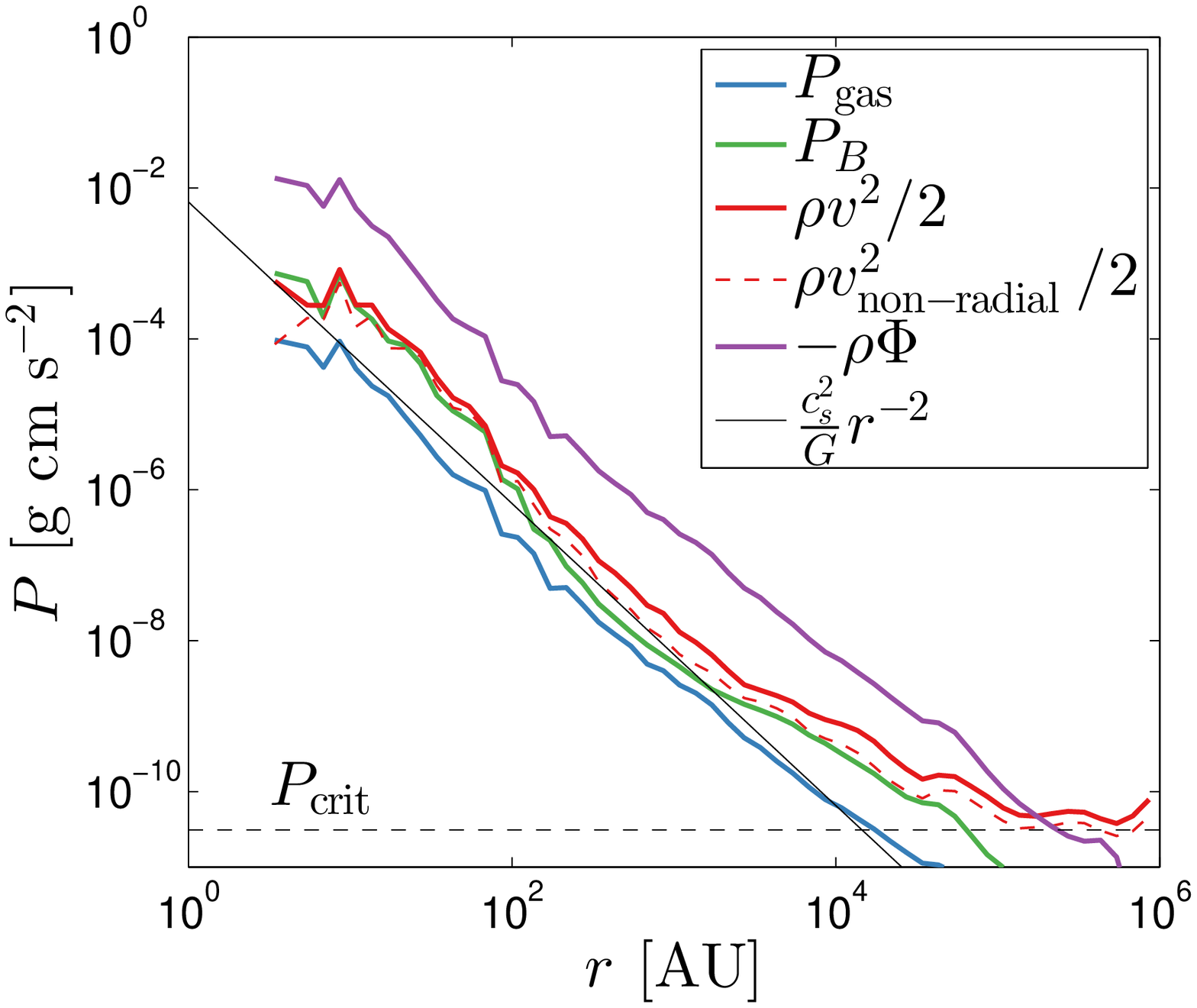} &
\includegraphics[width=0.45\textwidth]{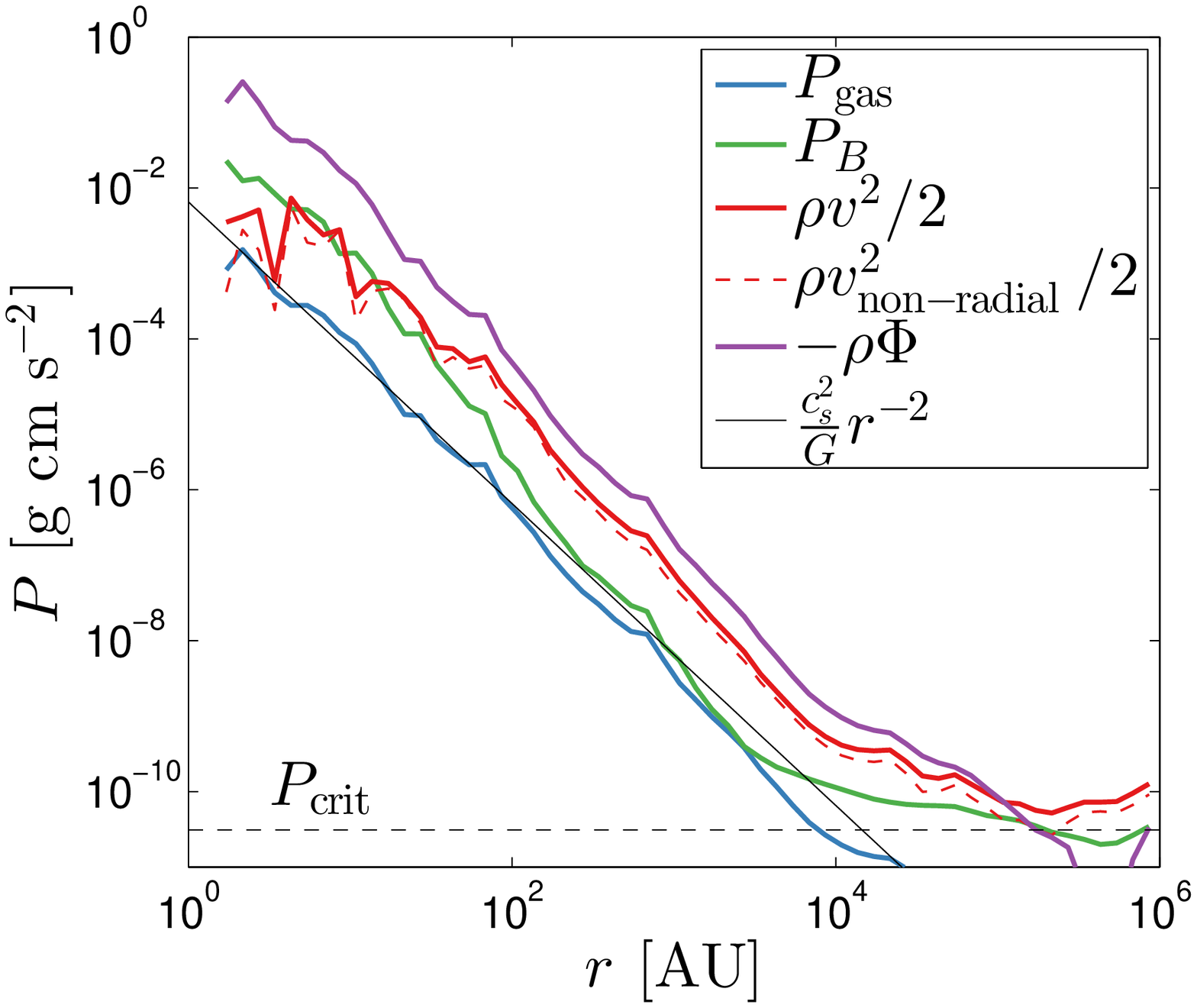} \\
& \\
{\Large $\MA0 = 1.2$} & 
{\Large $\MA0 = 0.35$} \\
\includegraphics[width=0.45\textwidth]{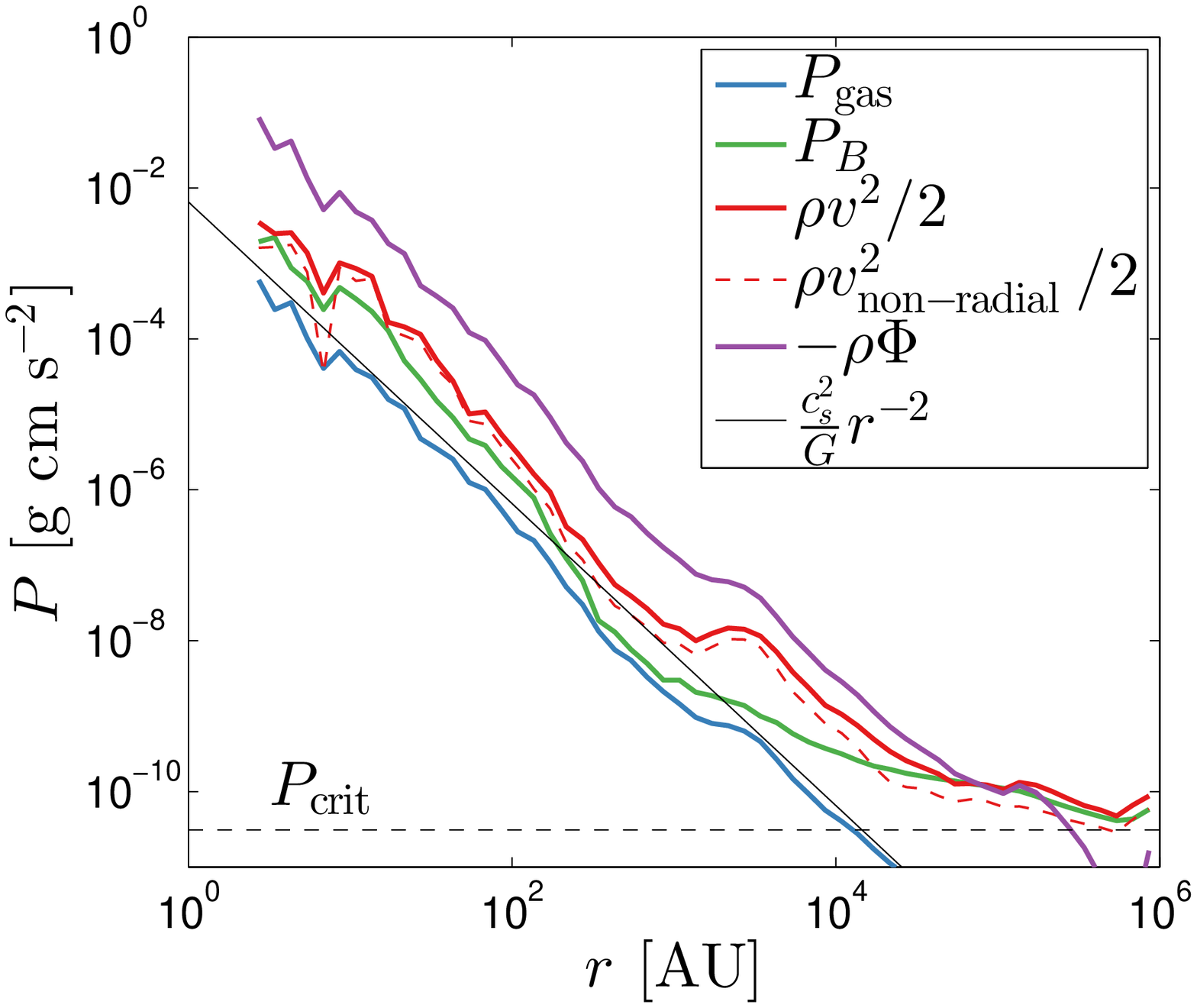} &
\includegraphics[width=0.45\textwidth]{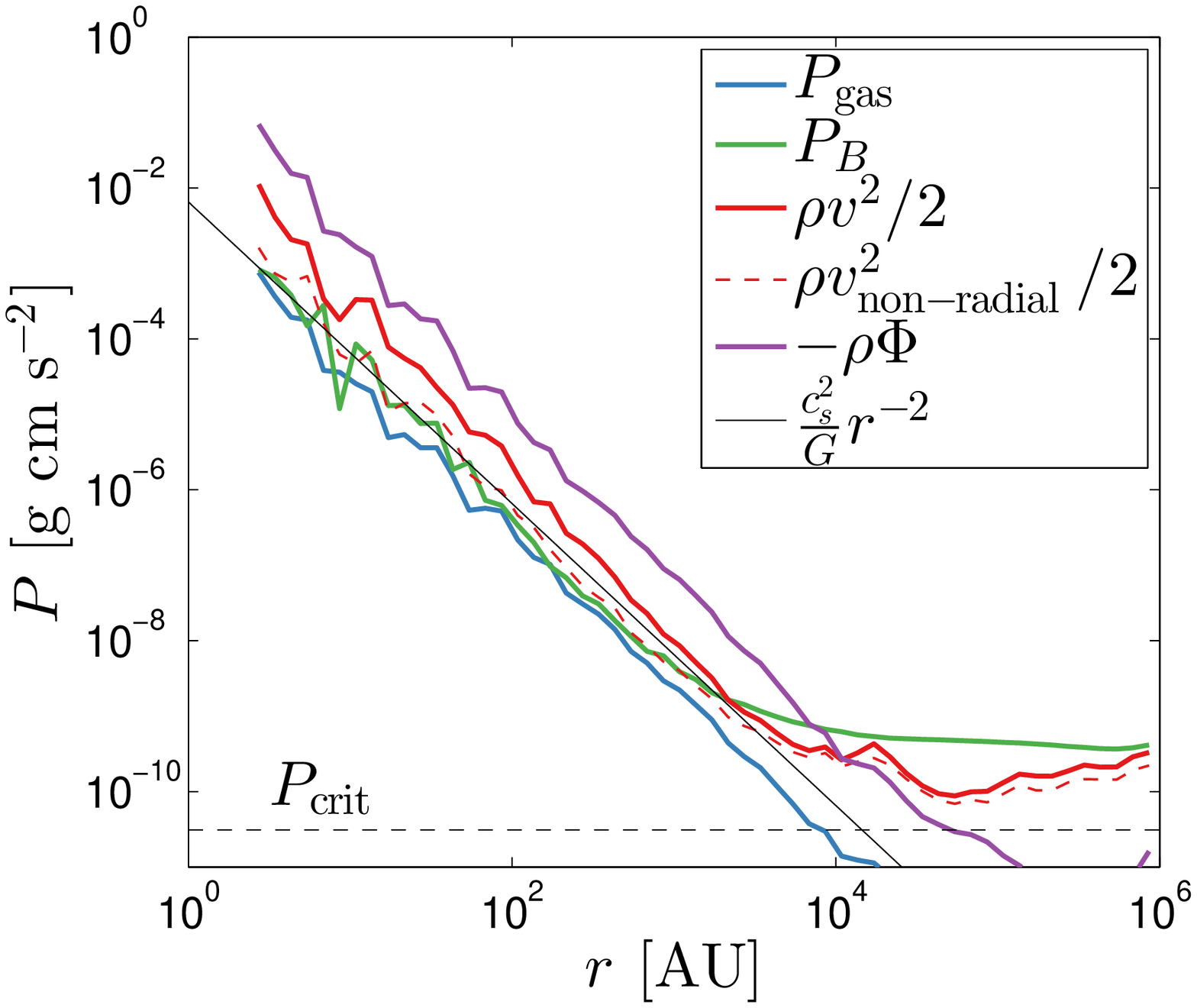}
\end{tabular}
\end{center}
\caption{Radial profiles of the gas pressure, magnetic pressure, kinetic energy density (also its non-radial component), and gravitational potential energy density for the four simulations. 
These profiles have been calculated using volume-weighted averages of gas cells contributing to radial bins. Also shown is the $\rho(r) = c_{\rm s}^2G^{-1}r^{-2}$ scaling, and the critical pressure $P_{\rm crit} = \rho_{\rm crit} c_{\rm s}^2$.}
\label{fig:sim_profile}
\end{figure*}

\section{Projections}
\label{sec:proj}
Fig.~\ref{fig:sim_proj} shows density-averaged line-of-sight magnetic field and column densities, zooming in on a $3000~{\rm AU}\times3000~{\rm AU}$ region of the densest collapsed core in each simulation (projections for the entire simulation domain are shown in Fig.~\ref{fig:sim_box}). The mean-field points in the horizontal ($x$) direction. The strong-field case ($\MA0=0.35$) shows highly-elongated structure (perpendicular to the mean-field direction) and a classical hourglass magnetic field morphology aligned with the mean-field direction. The weak-field simulations ($\MA0>1$) show more chaotic magnetic field morphology dictated by turbulence, with pinches and clumps. 
Fig.~\ref{fig:meanField} shows the deviation in the mean magnetic field ($\langle\Delta \theta_{\textrm{mean-field}}\rangle$) from its large (pc) scale direction as a function of scale: alignment becomes poorer with weaker mean-field strengths. The degree of alignment can remain tight to the smallest scales in the strong mean-field regime.
Fig.~\ref{fig:meanField} also shows a related quantity: the (volume) fraction of gas within a sphere of radius $r$ (the length scale) that has a magnetic field aligned within $30$~degrees of the large-scale mean-field value ($f_{\Delta \theta_{\textrm{mean-field}}<30^\circ}$).

\section{Density-Scaling of the Magnetic Field}
\label{sec:Brho}
The relationship between density and magnetic field is important for theoretical models and observations.
 $B\propto\rho^{2/3}$ is predicted for collapsing clouds with weak magnetic fields (i.e., gravitational energy dominating magnetic energy).
 The same scaling is also predicted for flux-frozen, isotropic collapse in general.
\cite{Lazarian2012} interpreted the observed $B\propto\rho^{2/3}$ from Zeeman measurements as a signature of reconnection diffusion being most efficient at lower densities (i.e. before collapse proceeds) in the weak field limit.  
In contrast, theoretical predictions for anisotropic contraction models, often expected in the strong field limit, predict a weaker scaling: $B\propto\rho^{1/2}$ \citep{2015MNRAS.451.4384T}.
 
We test the density magnetic field scaling relationship using our high dynamic range \textsc{Arepo} simulations. We plot the occupation of the gas in the density/magnetic field phase diagram for our simulations in Fig.~\ref{fig:sim_Brho}. The collapse occurs where the gas is compressed by turbulence above the critical density $\rho_{\rm crit} = \langle \rho \rangle \mathcal{M}_{\rm s}^2/3$ which is defined as the density at which the background level of turbulent pressure is sub-dominant to the gas pressure \citep{2005ApJ...630..250K,2015MNRAS.452.2500L}. That is, the critical pressure is obtained by equating $P_{\rm turb} = \frac{1}{3} \langle \rho \rangle v_{\rm rms}^2$ with $P_{\rm gas} = \rho c_{\rm s}^2$. Note this is not a strict condition for collapse, but is a necessary requirement.
Fig.~\ref{fig:sim_Brho} shows a moving-average of the $B$-$\rho$ correlation (yellow). The best-fit slope with $95$~per~cent confidence intervals to relation at $\rho>500\rho_{\rm crit}$ are listed in Table~\ref{tbl:sims}.
We find that when the mean-field is subdominant ($\MA0>1$), the collapse follows an approximately isotropic collapse, which is characterized by $B\propto\rho^{2/3}$. But in the strong-field case ($\MA0=0.35$) the collapse is anisotropic with $B\propto\rho^{1/2}$ and $\beta\approx 1$.

In Fig.~\ref{fig:sim_Brho} we also show inset-plots of the probability distribution function (PDF) of the gas density, which exhibits a log-normal distribution (indicative of turbulence, see \citealt{vazquez1997,burkhart2009,federrath2008}) and a $\rho^{-3/2}$ powerlaw tail (indicative of gravitational collapse see, \cite{kritsuk2007,2012ApJ...750...13C}). 
We find the critical density corresponds to the transition point between lognormal (diffuse turbulent gas) and power-law (self-gravitating) PDFs in these \textsc{Arepo} simulations, as observed also in \cite{2015MNRAS.452.2500L}. \cite{2005ApJ...630..250K} showed that the critical density was the condition for a clump to be self-gravitating.
Recently \cite{burkhart2016} worked out an analytic theory for the transition point between lognormal to powerlaw (which is termed the `post-shock density' in that work).

\section{Collapse profiles}
\label{sec:prof}
Fig.~\ref{fig:sim_profile} shows the volume-weighted core profiles from our simulations. The gas pressure, and hence density (since $P=\rho c_{\rm s}^2$, $c_{\rm s}$ a constant), follow a $r^{-2}$ power-law, as predicted by the simple analytic theory of isothermal spherical collapse which does not consider turbulence or the magnetic field \citep{1969MNRAS.145..271L,1977ApJ...214..488S}. This scaling is seen even in our strong magnetic field simulation. 
This allows us to define a relation between the density in the core and physical scale as: $\rho(r) = c_{\rm s}^2G^{-1}r^{-2}$. Note that the normalization (multiplicative prefactor) of this relation in our simulations is very close to unity, which is different from the non-magnetic, non-turbulent singular isothermal sphere, which has normalization $1/(2\pi)\simeq 0.16$.
Fits to the normalizations for our four (strong$\to$weak) simulations are: $0.46$, $0.60$, $0.83$, $0.58$ (95~per~cent errors to the fits are $\pm0.1$).
The normalization is somewhat close to the similarity solutions of spherically symmetric collapse solved numerically by \cite{1969MNRAS.145..457P,1969MNRAS.145..271L}, which have normalization $8.86/(4\pi)\simeq 0.71$.
The core profiles extend all the way out to the critical pressure value $P_{\rm crit} = \rho_{\rm crit} c_{\rm s}^2$ (beyond this the large-scale turbulent pressure exceeds the gas pressure; i.e., the size of a thermally supported core is about the sonic length). In all cases, we find the outer part ($\sim 10^4~{\rm AU}$) of the profile has a plasma-beta of $\beta\sim 1$ (note the ratio of blue and green lines in the Figure gives $\beta$). 
This is to be expected for $\MA0>1$ since the local Alfv\'en Mach number in the turbulence 
is of order unity, so $P_B\sim \rho \sigma_{\rm nt}^2 \sim \rho c_{\rm s}^2$ at the sonic length.
For $\MA0=0.35$, the turbulence does not significantly affect the field, so $B$ remains nearly constant until the gas pressure grows to the point
that it is comparable to the field pressure; thereafter, the collapse proceeds anisotropically with $B\propto\rho^{1/2}$, thus $\beta$ remains close to unity all the way to the center of the core. However, if the mean magnetic field is subdominant to the kinetic energy density ($\MA0>1$) then the magnetic field grows faster towards the center (since $B\propto \rho^{2/3}$). 
We also plot the kinetic energy density profile (and its non-radial component), which dominates over the gas pressure. 
Our cores have not formed any disks on these scales (there is no evidence of a Keplerian rotation profile); the non-radial component of the kinetic energy density originates from the large-scale turbulent motions.
In weak mean-field cases ($\MA0>1$), the magnetic field, growing faster than the other quantities towards the center, is in near equipartition with the kinetic energy density at the center. The collapse is in approximate Virial equilibrium, which is an assumption used in the turbulent core model \citep{1997ApJ...476..750M,2003ApJ...585..850M} and in the turbulent collapse model of \cite{2015ApJ...804...44M}. It should be noted these latter authors considered cores that are dominated by turbulent pressure rather than thermal pressure. For such cores, \cite{2003ApJ...585..850M} found an $r^{-3/2}$ profile from observations of star-forming regions, and \cite{2015ApJ...804...44M} inferred such a profile in the vicinity of a protostar from theory.

The normalized mass-to-flux ratios in our cores have evolved from the initial value of the box to:
 $\mu_{\Phi,0}: 80 \to 12.7$, $8 \to 16.5$, $2.7 \to 12.1$, $0.8 \to 5.8$ at a core-size of $10^4~{\rm AU}$. This is consistent with reconnection diffusion in our simulations. The mass-to-flux ratio cannot increase under flux-frozen, ideal-MHD conditions. Our strong-field simulation is originally subcritical $\mu_{\Phi,0} = 0.8$, and the simulation should not allow collapse unless reconnection diffusion is present, which leads to a mass-to-flux ratio above unity. In the calculation of the mass-to-flux ratio, the volume-averaged magnetic field in a spherical region of diameter $10^4~{\rm AU}$ is used.
We note that the transition from sub-critical ISM clouds to super-critical cores is observed via 21-cm, OH and CN  Zeeman observations and that no sub-critical cores are observed in the sample of \cite{2010ApJ...725..466C}.

Reconnection in our numerical simulations is enabled by numerical resistivity, but in this fast turbulent reconnection diffusion regime the simulation mimics the actual physical process because the reconnection rate is independent of the strength of the resistivity \citep{2009ApJ...700...63K}. Previous studies have found mass-to-flux ratios numerically converged with resolution \citep{2015MNRAS.452.2500L}, supporting the idea that a simulation with resolution such as ours captures the physical process accurately.

\section{Discussion}
\label{sec:disc}
We have presented basic scaling relationships of density and magnetic field, and radial profiles, obtained from novel high-resolution simulations, important for both theoretical models and observations of star formation. 
In a sense, the simulations advance the simple spherical, self-similar isothermal analytic collapse model \citep{1969MNRAS.145..271L,1977ApJ...214..488S}, with the inclusion of full 3D effects of magneto-turbulence. The classic $r^{-2}$ radially-averaged pressure profile scaling is still recovered in all cases, with similar normalization across simulations. The same $r^{-2}$ scaling is seen in simulations of turbulence-dominated, nearly isothermal atomic cooling halos in \cite{2015MNRAS.446.2380B}.
Our simulations do not show inside-out collapse, which would lead to a $r^{-3/2}$ profile \citep{1977ApJ...214..488S}.
We point out that we stop the simulations before we would expect stars to form, meaning, our cores are pre-stellar.
Thus we do not yet see a $r^{-3/2}$ scaling in the vicinity of the star (inside the sphere of influence, after the star forms), as predicted by turbulent models for star formation in compact massive clouds from non-hydrostatic initial conditions of \cite{2015ApJ...804...44M}, and as observed in the simulation setup of \cite{2015arXiv150905910M}.

Collapse in the turbulent medium occurs where the gas pressure exceeds the background turbulent pressure; i.e., above the critical density $\rho_{\rm crit}$.
At these outer-scales of collapse, we find the plasma beta is always near unity regardless of mean-field strength, indicating equipartition between the magnetic and gas pressures in the turbulent environment. 
A plasma beta near unity occurs in the $\MA0>1$ simulations because in the turbulent (uncollapsed) environment the magnetic field and density are not well correlated, instead the local Alfv\'en Mach number in the turbulence  is of order unity, so at the sonic length $P_B\sim \rho\sigma_{\rm nt}^2 \sim \rho c_{\rm s}^2$.
In the strong-field regime, the field strength is not affected much by turbulence, so $B$ remains nearly constant until the gas pressure grows to the point that it is comparable to the field pressure (which happens at the tail of the log-normal distribution of density which develops from turbulence).

Anisotropic collapse with $B\propto\rho^{1/2}$ is exhibited in the sub-Alfv\'enic simulation, and we form cores with a classic hourglass-like magnetic field morphology, similar to NGC 6334 \citep{2015Natur.520..518L}. 
But when the mean-field is weak ($\MA0>1$), the collapse is spherical and hence $B\propto\rho^{2/3}$, which means that the plasma beta decreases toward the center. Interestingly, due to the $B\propto\rho^{2/3}$ scaling, the weak-field case actually allows for a stronger magnetic field in the core than in the strong-field ($\MA0=0.35$) case, where the magnetic field does not rise as fast with density. Constant plasma beta self-similarity in the collapse is broken in the weak-field case. There are two clear regimes of magnetic field evolution we observe, but we do not mean to conclude or imply that the transition is very sharp, occurring exactly at $\MA0=1$.

Even if the initial large-scale mean-field is weak, turbulence amplifies it considerably prior to collapse. We attribute the decrease in plasma beta in the super-Alfv\'enic simulations as a result of the small-scale turbulent dynamo process.
Our simulations show the initial plasma beta evolves as:
$\langle\beta\rangle: 0.0025\to 0.0025$, $0.028\to 0.025$, $0.25\to 0.064$, $25\to 3.2$; i.e., the initial plasma beta shrinks in the super-Alfv\'enic cases. The average plasma beta is not increased significantly in the strong mean magnetic field run by turbulence.
We note all simulations are driven with the same amplitude, with sonic Mach number $10$.
The small-scale turbulent dynamo is very efficient in amplifying the magnetic field and may in fact be more efficient in the super-Alfv\'enic limit \citep{cho2008}. 

As the turbulent simulations evolve and collapse, magnetic field is generated by the stretching and twisting of magnetic field lines and through flux conservation. However, turbulence also acts to remove field lines from collapsing regions via reconnection diffusion and the magnetic field changes its topology in just an eddy turn over time \citep{Lazarian1999,Vishniac2003,Lazarian2004}. Once rapid collapse begins, reconnection diffusion, which depends only on the properties of turbulence/turbulence amplitude, removes magnetic field from the contracting clouds in competition with the amplification the field experiences due to contraction and dynamo processes. Future studies will determine how fast reconnection diffusion is compared to the dynamo process in such simulations in order to quantify the competition between flux removal and field amplification. Our simulations do show evidence for reconnection diffusion through the increase of the mass-to-flux ratios in the cores (\S~\ref{sec:prof}).

We note that our results should be as applicable to observations as they are to theoretical predictions. 
In all the simulations, we transformed micro-Gauss level large-scale fields into milli-Gauss level core-scale fields, in agreement with  various observational estimates of field-strengths \citep{2010ApJ...725..466C, 2006Sci...313..812G,2009Sci...324.1408G,2013ApJ...769L..15S, 2016ApJ...820...38H}. Despite different scaling relationships and core morphologies between the sub- and super-Alfv\'enic simulations, all our simulations reproduce \textit{similar} magnetic field strengths on core scales. 
The star formation process is imprinted with the self-similar nature of turbulence and gravitational collapse despite different initial environmental conditions.  
What sets formation with different mean-field strengths apart is the orientation of the mean-field on various length scales relative to the large-scale value (Fig.~\ref{fig:meanField}). The orientation remains well-preserved to $100~{\rm AU}$ scales (ALMA resolution) \textit{only} in our strong-field simulation ($\MA0=0.35$). If the magnetic field is moderately strong ($\MA0 \gtrsim 1$) then the mean-field only remains aligned to scales of $10000~{\rm AU}$ (resolution of the Combined Array for Research in Millimeter-wave Astronomy; CARMA). 
More precisely, the field remains aligned to less than $30$~degrees from its large-scale value down to scales of
$10^{2.0}~{\rm AU}$,  $10^{3.7}~{\rm AU}$, $10^{4.0}~{\rm AU}$, $10^{5.5}~{\rm AU}$ for our four (strong$\to$weak) simulations, as listed in Table~\ref{tbl:sims}.

Fig.~\ref{fig:meanField} shows the angle of magnetic field alignment averaged over the core as a function of scale, as well as the fraction of gas in a given radius that is aligned within $30$~degrees of the large-scale mean-field. Table~\ref{tbl:sims} also lists the fraction of gas aligned within $30$~degrees of the large-scale mean-field on a scale of 10000 AU. The field is strongly correlated in the large-scale strong-field case and virtually uncorrelated if the large-scale mean-field is very weak. The correlation is in between for the intermediate mean-field cases.  (Note the curves for $\MA0=1.2$ and $\MA0=3.5$ cross each other: there is a certain amount of variance to be expected in the correlation as function of $\MA0$ for a large population of cores, which we do not have the statistics to probe in our current simulation because we focus on resolving single cores.) Observationally, \cite{2009ApJ...704..891L} found a strong correlation between orientation on the 10000 AU scales and large scales, whereas \cite{2014ApJS..213...13H} did not for smaller scales. However, there is some ambiguity on the latter since it could be confused by toroidal wrapping in a disk, and CARMA also only provides a few independent polarization vectors, not enough to determine the mean-field direction with robust statistics. This makes ALMA paramount for the study of the importance of magnetic fields on scales less than 10000 AU.

We note that the magnetic field, which is found to be always strong in the core regardless of the large-scale value (local Alfv\'enic Mach number is order unity in the core), has a significant effect on the angular momentum of the accreting gas and thus on the properties of circumstellar, including protoplanetary, disks, that are to form.
Future studies resolving the core collapse beyond the isothermal collapse stage will help explain the ultimate fate of the protostar. The present study serves as providing useful initial conditions to this problem.
An interest of study is the `magnetic braking catastrophe', where magnetic fields can in theory prevent the formation of circumstellar discs around young stars (e.g., see \cite{2016MNRAS.457.1037W} and references therein). Our chaotic magnetic field morphologies in the weak mean-field simulations perhaps could be one way around the magnetic braking catastrophe, through the effects of flux loss via reconnection diffusion.  The influence of turbulence in reducing magnetic torques has been examined in a number of papers \citep{2012MNRAS.423L..40S,2013A&A...554A..17J}.

Our $B\propto \rho^{2/3}$ scaling relation for the weak-field simulations is in agreement with the Zeeman observations of diffuse and molecular clouds of \cite{2010ApJ...725..466C}, which see $\langle B\rangle_M\propto \langle \rho\rangle^{0.65}$.
We note, however, that the relationship between $B$ and $\rho$ in our simulations, as indicated by the moving-average fits (yellow lines) in Fig.~\ref{fig:sim_Brho}, shows the slope of the correlation transitioning from a flatter-than-$2/3$ value in the turbulent medium to the $2/3$ slope in the collapsed cores. The tight scaling of $B\propto \rho^{2/3}$ is seen on scales of $r=10^3$-$10$~AU (i.e., $\rho>500\rho_{\rm crit}$).
The slope is actually closer to $B\propto \rho^{0.5}$ in the collapsed-regions with $\rho_{\rm crit}<\rho<500\rho_{\rm crit}$, similar to the simulations of \cite{2012ApJ...750...13C}.
That is, we find that the slope is actually shallower than $2/3$ on scales of $r=10^3$-$10^4$~AU due to the transitioning between the turbulent medium and cores. These scales are more relevant for the clump sizes observed in \cite{2010ApJ...725..466C} for deducing the observational scaling relation.
\cite{2015MNRAS.452.2500L} carried out a more thorough analysis of reproducing the \cite{2010ApJ...725..466C} relation adding in all of the observational effects, including convolution with a beam size of $\sim 5000$~AU. Taking density-averaged (as measured by Zeeman observations) as opposed to volume-averaged magnetic field strengths, \cite{2015MNRAS.452.2500L}
 found a value of $\alpha$ consistent with $0.65$ for a moderately strong
initial magnetic field ($\MA0=1$) but only marginally so for a weak initial field ($\MA0=10$).
 In contrast, the $B$--$\rho$ relation measured on a cell by cell basis on these scales has a flatter power. This demonstrates the importance of adding observational effects to simulation results when interpreting data. Similarly, \cite{2015MNRAS.452.2500L} show that in the calculation of the mass-to-flux ratio of clumps, the average area-weighted (i.e., flux/area) field is somewhat less than the \textit{observed} mass-averaged field, so that the actual mass to flux ratio is about $0.7$ times the value inferred from line-of-sight values, rather than (0.25-0.5) times the line-of-sight value, as is often used. Again this signifies the importance of putting simulation data through an observational pipeline.

Our simulations also demonstrate the importance of having a high-dynamic range calculation. The theoretically expected scaling of $B\propto \rho^{2/3}$ in magnetic field density phase-space is accurately resolved by being able to simulate scales of $r=10^3$-$10$~AU; i.e.,
convergence of the slope of the relation is achieved on the smallest scales; on larger scales the slope is flatter as it transitions to the uncollapsed background turbulent environment. 
In contrast, previous AMR simulations of this type have resolved down to scales of $120$~AU \cite{2012ApJ...750...13C} and $500$~AU \cite{2015MNRAS.452.2500L}. Our simulations are expected to have less numerical magnetic reconnection (which can remove magnetic flux and affect the scaling relation) as well as significantly reduced advection errors (which can be quite significant in supersonic flows) due to the quasi-Lagrangian nature of the moving-mesh formulation.


Comparing with the results of various observation of the ISM (e.g. \citealt{2010ApJ...725..466C} and \citealt{2015Natur.520..518L}), our simulations suggest that $\MA0\sim 1$ may be typical in many ISM regions (where `$\sim$' means the coefficient of the relation is unconstrained by an order of magnitude). The magnetic field is often observed to be coherent on large-scales and the density structure is filamentary (thus we do not expect $\MA0\gg 1$). Our simulations suggest that the existing observations for star formation provide evidence that the collapse occurs in \textit{both} the regimes $\MA0\gtrsim 1$ and $\MA0\lesssim 1$. The former is supported by $\langle B\rangle_M\propto \langle \rho\rangle^{0.65}$ Zeeman splitting observations of dense molecular cloud clumps
\citep{2010ApJ...725..466C} while the latter is supported by multiple observations of NGC 6334, which show self-similar, aligned magnetic field structure from $100$~pc to $0.01$~pc scales with $B\propto\rho^{0.41}$ \citep{2015Natur.520..518L}. This is assuming that the determination of the slope of the magnetic field density relation in \cite{2015Natur.520..518L} is robust to the indirect methods used. Note, we do not claim that the transition between regimes occurs as a \textit{sharp} transition at  $\MA0=1$, just that two different regimes of magnetic field evolution (under the ideal MHD assumptions) exist.

Regions in the ISM with stronger magnetic fields may have suppressed star formation. This is seen in the strong-field box, which takes longer to collapse and form the first core because it is initially slightly sub-critical and requires turbulent reconnection to reach criticality to collapse under self-gravity. Even in the other simulations, the time of the collapse of the first core correlates weakly with magnetic field strength, because the magnetic field provides extra pressure support against collapse. This effect may have relevance for the star formation in the galactic center, where star formation is observed to be suppressed and the magnetic field is strong: $500~\mu{\rm G}$--$5{\rm mG}$ 
\citep{2016A&A...591A..19P,2016arXiv161107022K}.

We discuss the caveat that turbulent collapsing simulations such as ours generate hundreds of cores over a single free-fall timescale. However, our simulation strategy was to resolve features on $<100~{\rm AU}$ scales (relevant for ALMA observations) and therefore we did not impose any minimum spatial resolution or sink particles. This comes with a trade-off, however: the timestep of the simulation decreases exponentially as the first core forms, thus we are only able to resolve (at $<100~{\rm AU}$) the first-prestellar core that collapses, which happens early on, at $<0.4t_{\rm ff}$. We are not able to say much about a large statistical sample of collapsed cores in the present work, because most cores have not yet collapsed to such spatial scales. Importantly, however, the collapsed cores we investigate are not outliers from a powerlaw distribution of masses, as they all have similar masses. (Their core masses are also not extreme compared to the masses of the other turbulently fragmented cores in the box which are at an earlier stage of collapse; see Appendix~\ref{sec:allcores}.). It is plausible that certain geometries (e.g. relative orientation between the magnetic field and angular momentum vector) would collapse first in a turbulent medium, so there could be other types of bias present in the types of cores that collapse first. We defer analysis of a statistical sample of cores to upcoming future work, in which we will modify our simulation strategy and trade-off resolution and the capture of the exact collapse process with the use of sink particles.

\section{Conclusions}
\label{sec:conc}
We have simulated the formation of pre-stellar cores in a supersonic, turbulent, magnetic interstellar medium under self-gravity, resolving the initial isothermal collapse phase down to a few AU, relevant for future ALMA polarization observations of Class 0 protostars. We studied the effects of the mean-field strength, and have arrived at the following main conclusions:

\begin{itemize}
\item If the turbulent kinetic energy density of the ISM dominates over the mean-field magnetic pressure, then collapse occurs approximately isotropically on small scales ($r<10^4~{\rm AU}$), with $B\propto\rho^{2/3}$. But in the case of strong large-scale magnetic field, the collapse is anisotropic with $B\propto\rho^{1/2}$.
\item On the larger scales of collapsed cores ($r>10^4~{\rm AU}$), the scaling of magnetic field with gas density (looking at it on a simulation cell-by-cell basis) is flatter, close to $B\propto\rho^{1/2}$ in all cases, as the gas transitions to the turbulent background medium. 
The results of \cite{2015MNRAS.452.2500L} show that if one instead looks at the scaling of the 
 \textit{density-averaged} magnetic field (convolved with the telescope beam) $\langle B\rangle_M$ with $\langle \rho\rangle$, the slope of the correlation is steeper, closer to $0.64$. \cite{2010ApJ...725..466C} observe cores have $\langle B\rangle_M\propto \langle \rho\rangle^{0.65}$ via Zeeman splitting measurements on scales of $r>10^4~{\rm AU}$. This highlights the importance of modeling observational biases when interpreting physical results and comparing observations with simulations.
\item Our simulations of collapsing pre-stellar cores all follow similar radially-averaged $r^{-2}$ profiles (as predicted by analytic spherical isothermal collapse model), regardless of mean-field strength. In the weak-field case, even though the collapse is isotropic, the $B\propto\rho^{2/3}$ scaling means the collapse is not self-similar in the sense that $\beta\equiv P_{\rm gas}/P_B$ decreases towards the center and the Alfv\'enic Mach number also drops with radius. These quantities are approximately constant in the case of strong-field collapse. Collapse occurs in approximate Virial equilibrium.
\item Regardless of cloud scale mean-field strength, the outer regions ($\sim 10^4~{\rm AU}$) of the cores have $\beta\sim 1$ and the magnetic field reaches equipartition with the thermal energy, which in turn is comparable to the average kinetic energy density in the simulation box.
\item The magnetic field in the center of the cores is actually slightly stronger (by about a factor $3$ on $100$~AU scales) if the large-scale mean-field is weaker than the turbulent kinetic energy density because of the $B\propto\rho^{2/3}$ scaling as opposed to $B\propto\rho^{1/2}$.
\item The mass-to-flux ratios in our cores increase by a factor of a few from the large-scale value, indicating fast reconnection diffusion. 
\item If the large (pc) scale mean-field is subdominant to the turbulent kinetic energy density ($\MA0>1$), then the magnetic field on $100$~AU core scales has uncorrelated direction with the mean-field (although may still be correlated on larger core scales $\sim 0.1$~pc). If the field is strong, the field direction remains correlated on all scales, and has a classic hourglass-like morphology. Correlation on intermediate scales ($10000$~AU) is a function of $\MA0$, ranging from strongly correlated ($\MA0=0.35$) to virtually no correlation ($\MA0=35$) as shown in Fig.~\ref{fig:meanField}. The fact that correlation extends to smaller scales as $\MA0$ decreases makes future ALMA observations very useful for constraining $\MA0$ of the ISM.
\item In terms of upcoming directions/opportunities in this subfield, observationally quantifying the scale at which the orientation correlation drops in a statistically significant sample of molecular cloud cores will be very useful in improving our knowledge of the ISM. 
\end{itemize}

\acknowledgments

P.M. is supported by the NASA Earth and Space Science Fellowship. 
B.B. is supported by the NASA Einstein postdoctoral fellowship.
The research of C.F.M. is supported in part by NSF grant AST-1211729 and NASA TCAN grant NNX-14AB52G.
The authors thank Chat Hull, Alex Lazarian and Pak-Shing Li for valuable discussions and reading of the manuscript. 
Computations were run on the Odyssey cluster supported by the FAS Division of Science, Research Computing Group at Harvard University. 
\section*{References}

\bibliography{mybibfile}

\appendix

\section{Turbulent fragmentation in simulations}
\label{sec:allcores}

Turbulent fragmentation leads to the collapse of hundreds of cores inside our simulation volume, as has been seen previously \citep{2011ApJ...731...59C,2012ApJ...750...13C,2015MNRAS.452.2500L}. In this paper, we have focused our simulation efforts on capturing the collapse of the first core that forms. Here we compare some of the properties of these cores to the larger samples of pre-stellar cores in our boxes, which are at an earlier stage of collapse (typically the core has only collapsed to scales of $10^3~{\rm AU}$). Cores are identified by selecting local maxima of the density field within a radius of $1500~{\rm AU}$.

\begin{figure*}
\begin{center}
\begin{tabular}{cc}
{\Large $\MA0 = 35$} &
{\Large $\MA0 = 3.5$} \\
\includegraphics[width=0.45\textwidth]{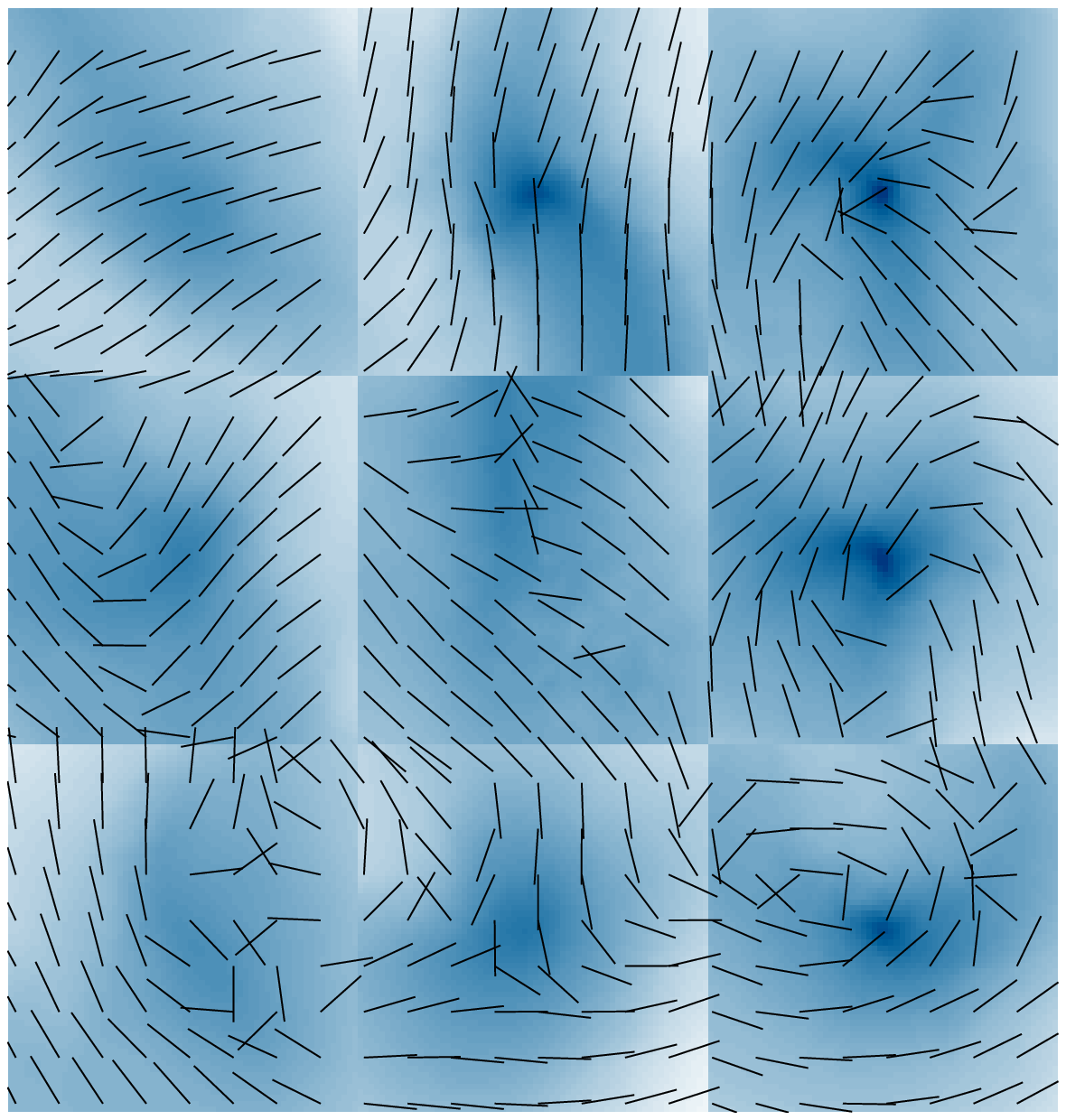} &
\includegraphics[width=0.45\textwidth]{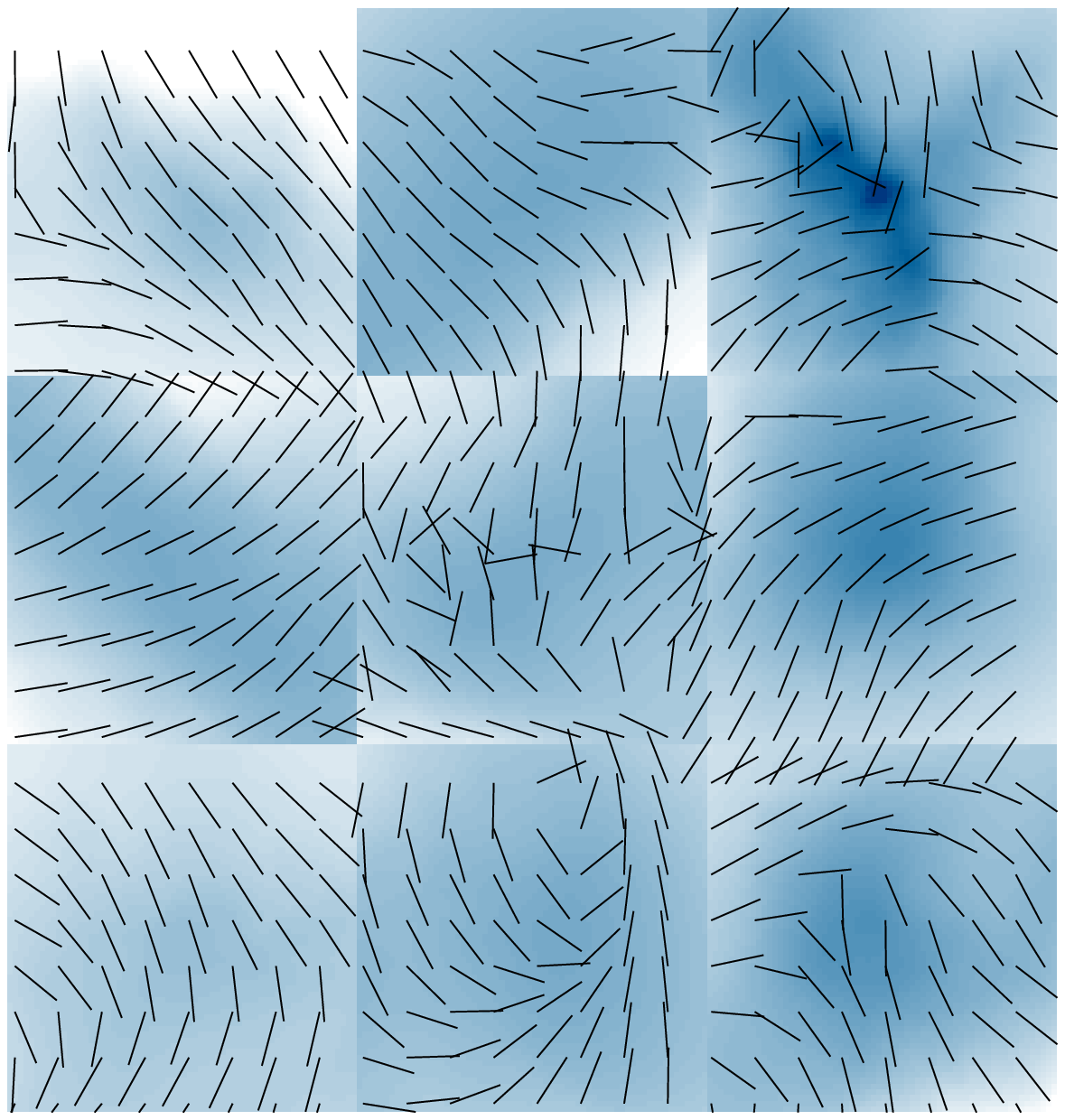} \\
& \\
{\Large $\MA0 = 1.2$} & 
{\Large $\MA0 = 0.35$} \\
\includegraphics[width=0.45\textwidth]{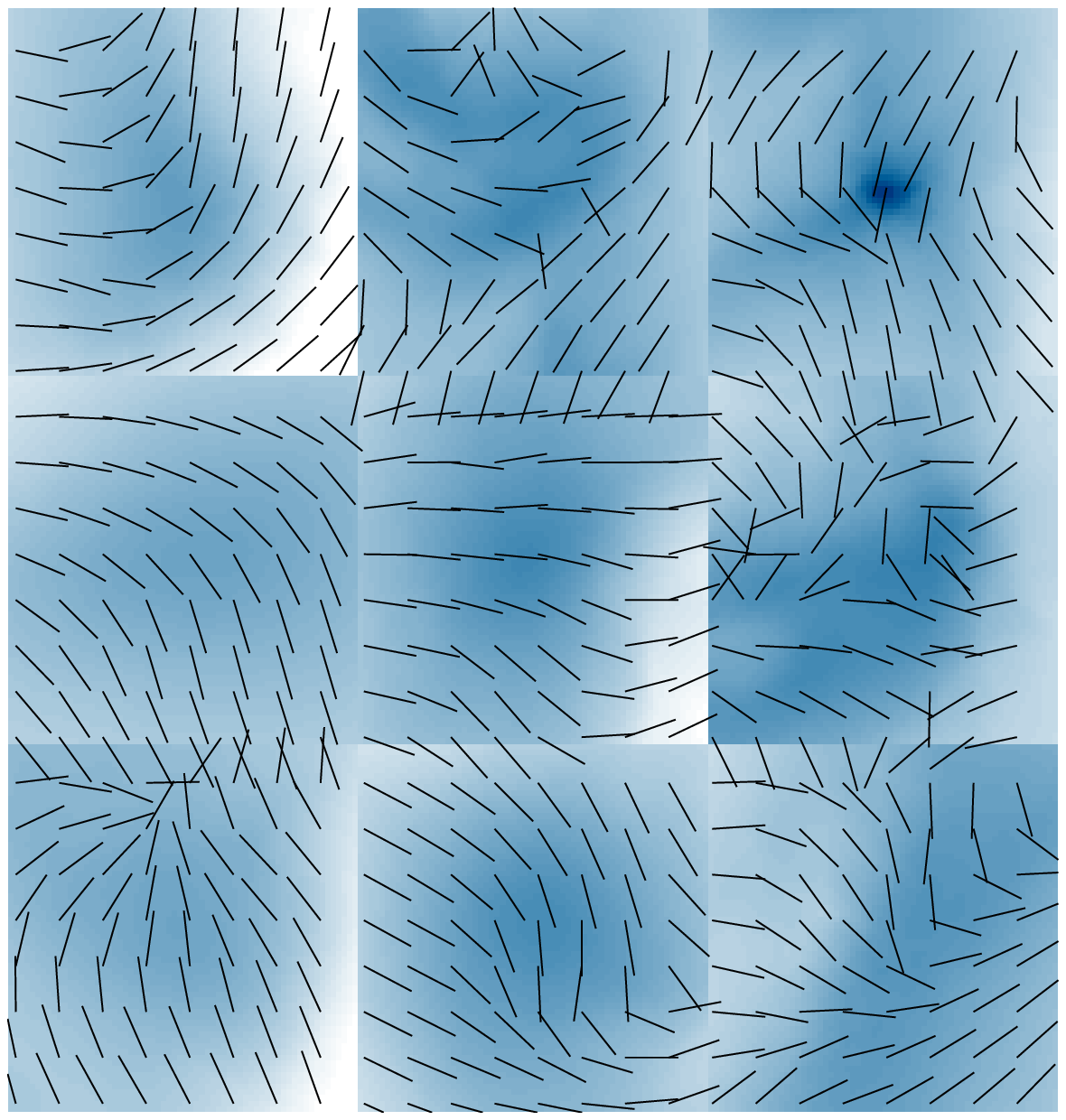} &
\includegraphics[width=0.45\textwidth]{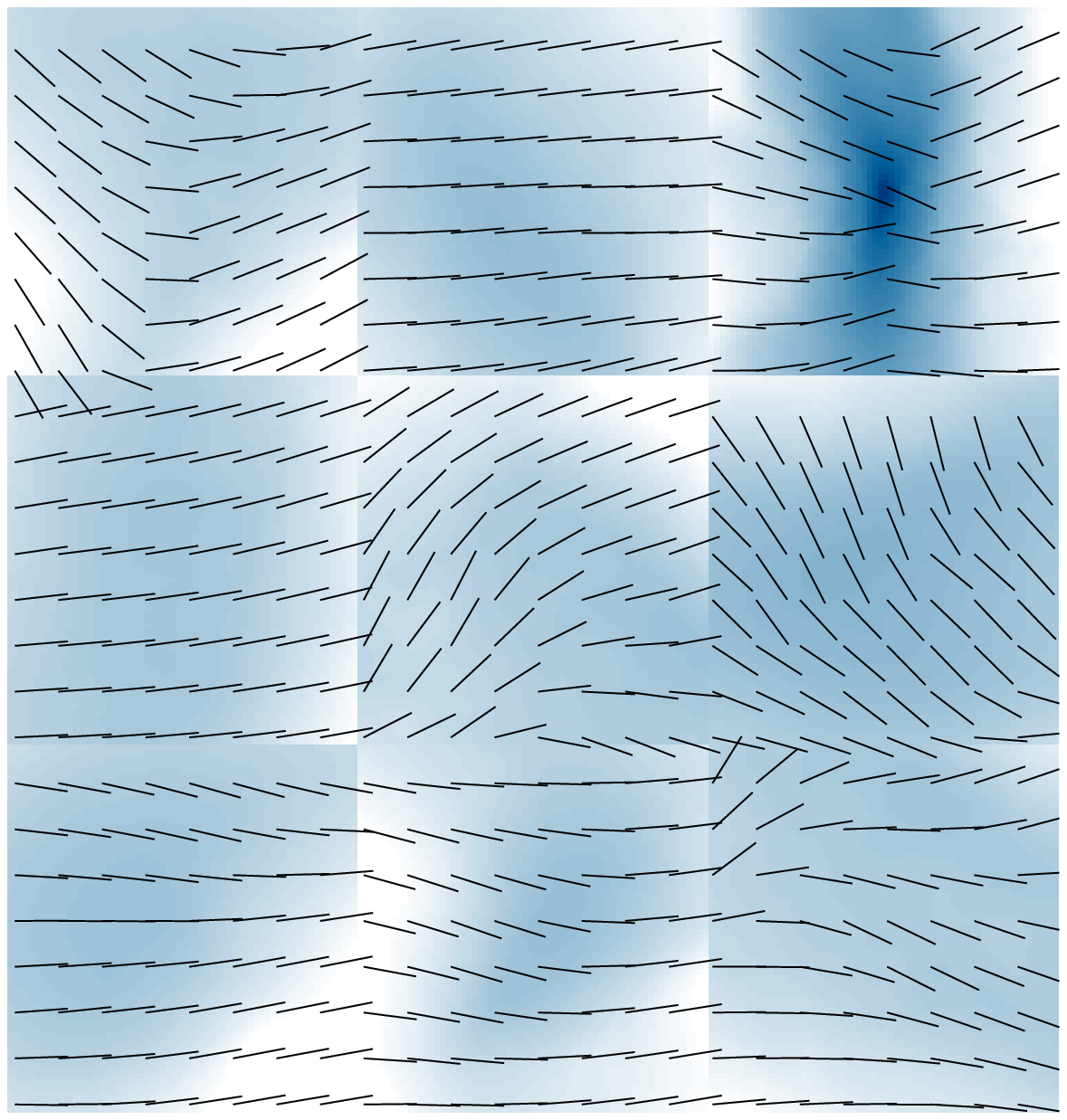}
\end{tabular}
\end{center}
\caption{Magnetic field and density projections of the $9$ most-collapsed cores in each of the simulations. Similar to Fig.~\ref{fig:sim_proj}. When the field is strong ($\mathcal{M}_{\rm A}=0.35$) the magnetic field aligns well with the mean-field and shows simple structure. In the other cases, there is evidence for chaotic field morphology even in some of the less-collapsed cores.}
\label{fig:allcores}
\end{figure*}

The analysis allows us to learn about how wide spread is the existence of hourglass figures.
Fig.~\ref{fig:allcores} shows the field morphologies for the $9$ most-collapsed cores in each of the simulations. 
We see that in the case of a strong magnetic field ($\mathcal{M}_{\rm A}=0.35$) the field lines align well with the mean-field direction. The morphologies are mostly linear on this scale, with some amount of pinching towards the core, and are likely to form hourglass shapes as the cores continue to collapse. But in the case of the simulations where the turbulent kinetic energy dominates the magnetic field, the morphologies can be quite chaotic on scales of $1000~{\rm AU}$, even in some of the less-collapsed cores. Therefore, the hourglass shape may be uncommon in such environments. The possibility of hourglass shapes is not excluded, however: a few cores show fairly linear morphologies with small amount of pinching towards the core center, which may end up evolving into an hourglass shape. 

\begin{figure*}
\begin{center}
\begin{tabular}{cc}
{\Large $\MA0 = 35$} &
{\Large $\MA0 = 3.5$} \\
\includegraphics[width=0.45\textwidth]{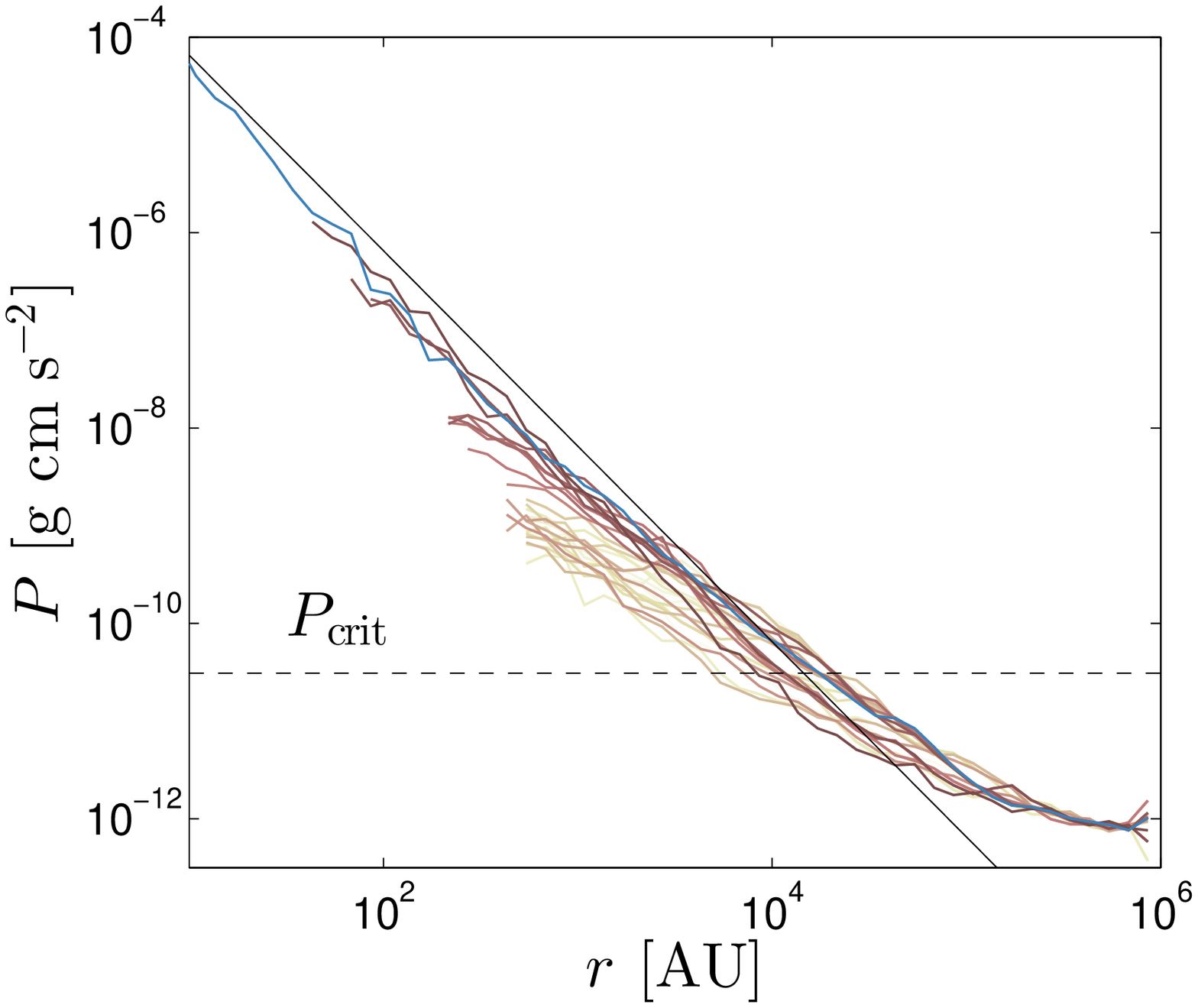} &
\includegraphics[width=0.45\textwidth]{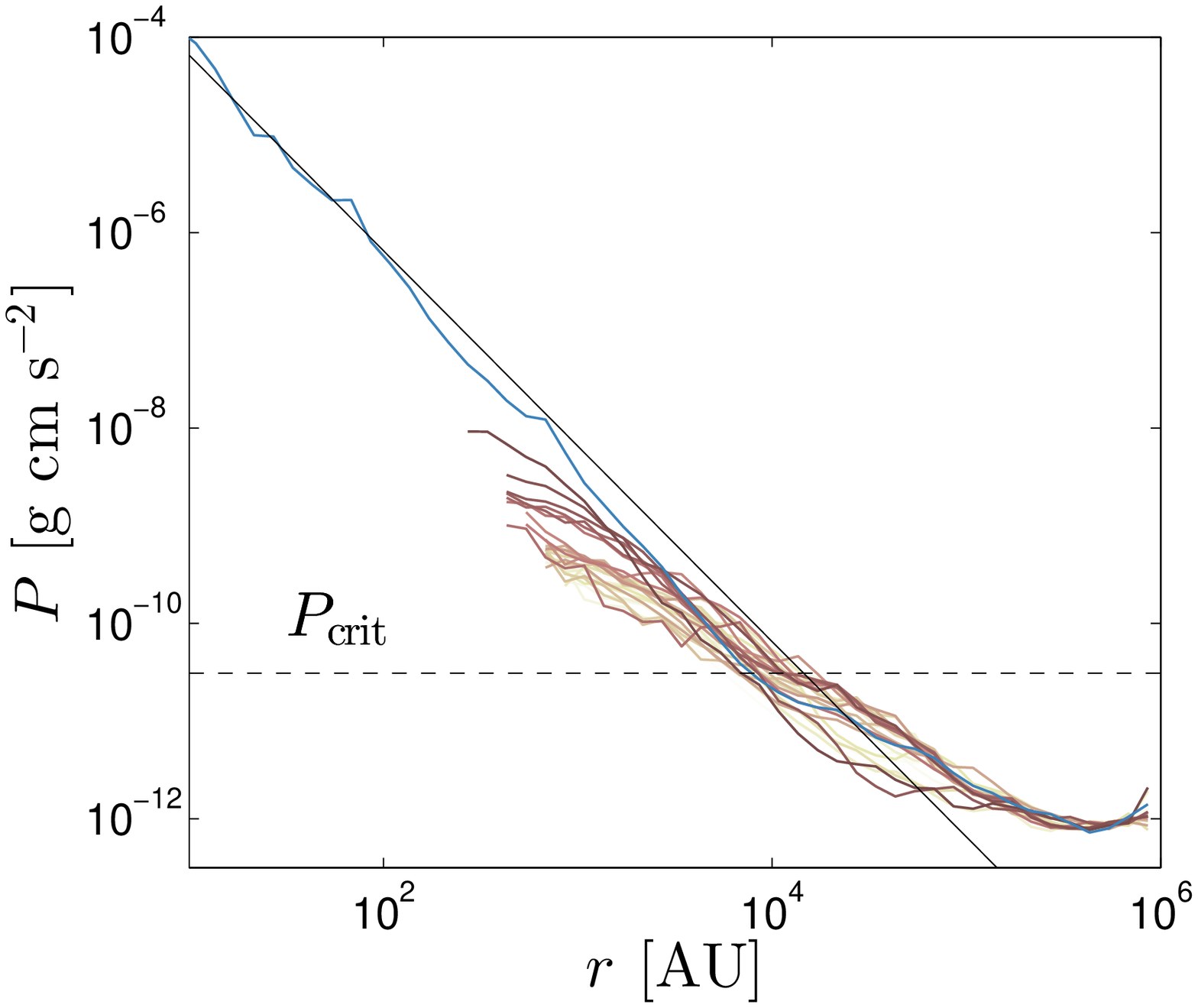} \\
& \\
{\Large $\MA0 = 1.2$} & 
{\Large $\MA0 = 0.35$} \\
\includegraphics[width=0.45\textwidth]{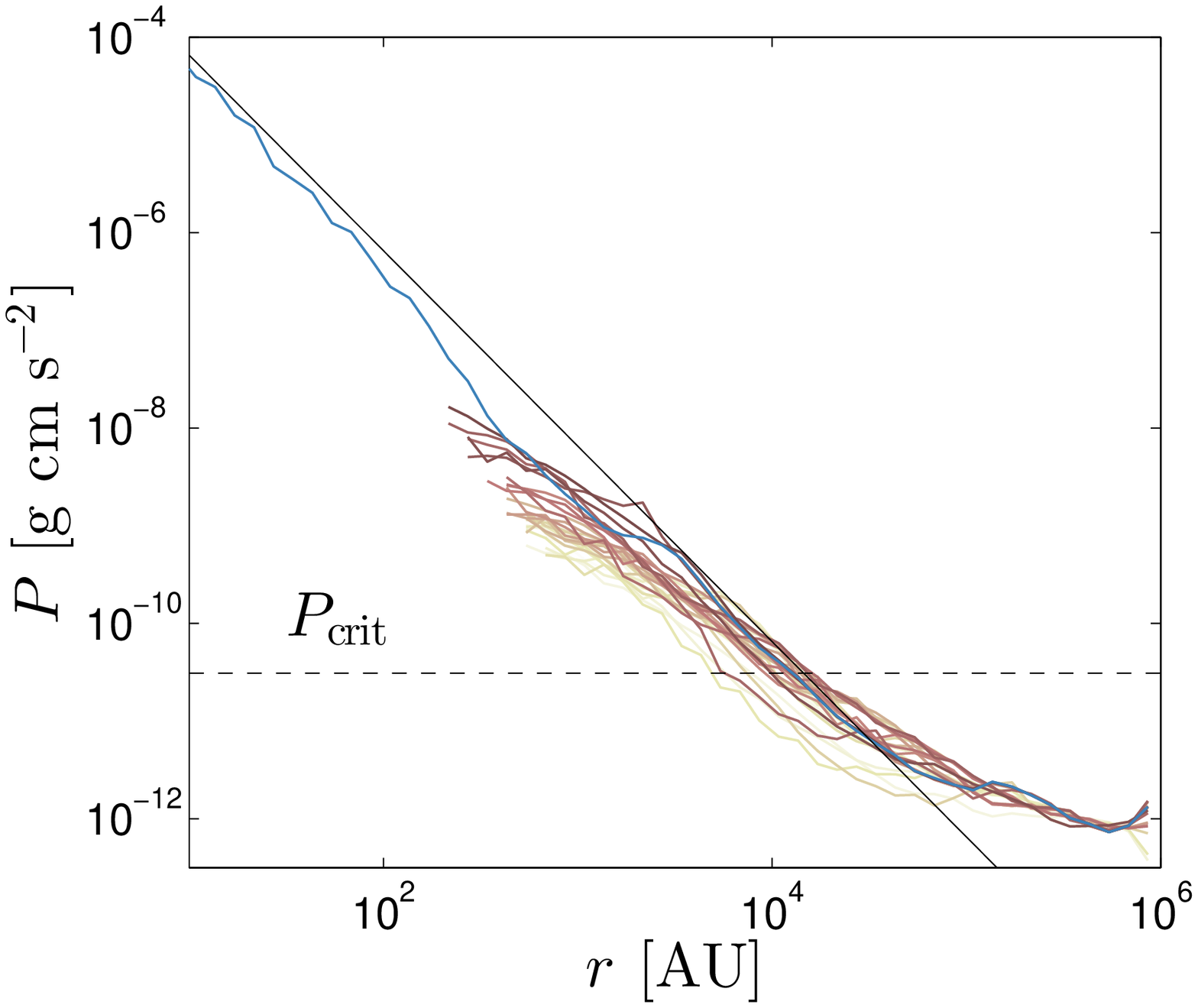} &
\includegraphics[width=0.45\textwidth]{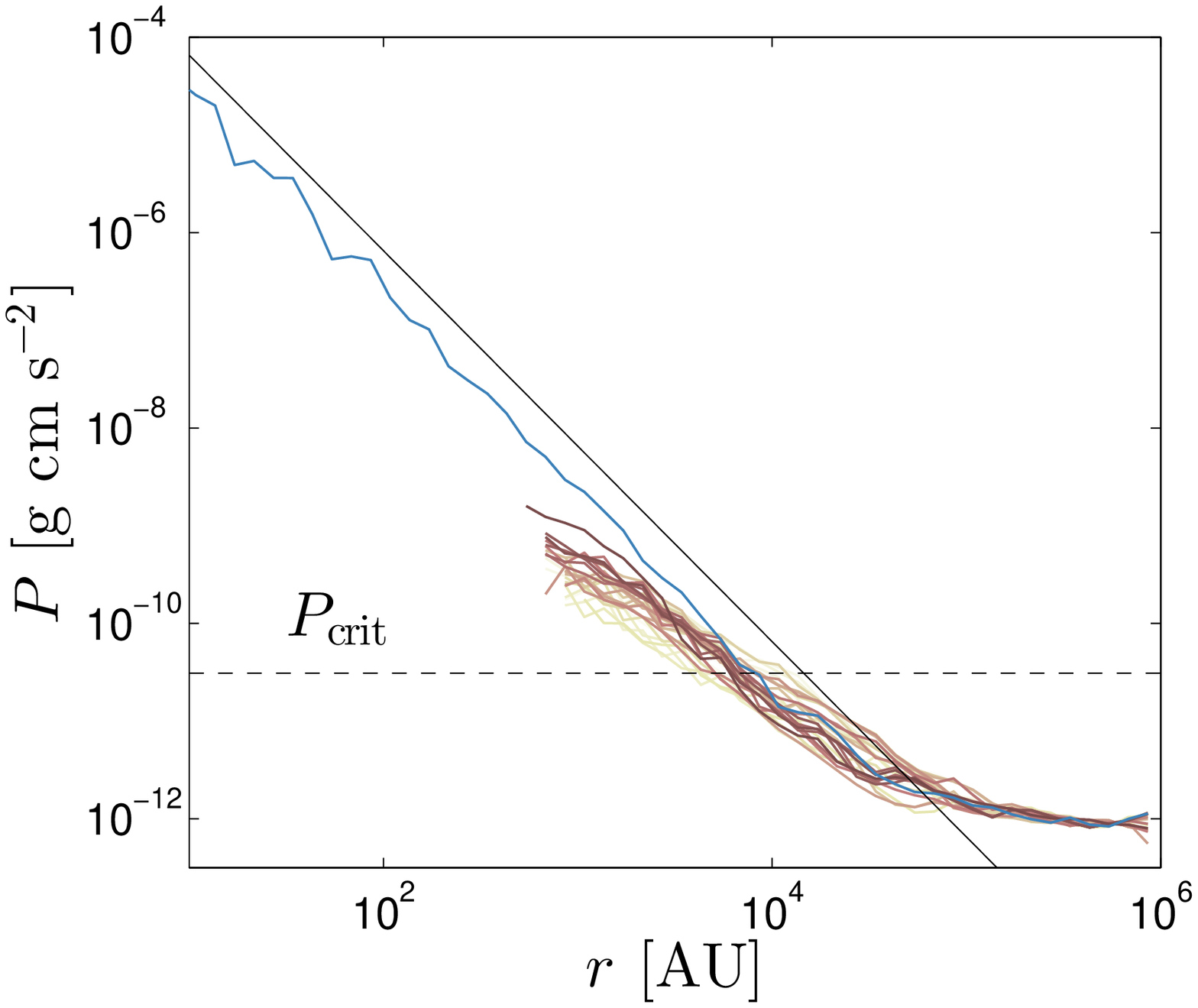}
\end{tabular}
\end{center}
\caption{Radial profiles of the density (plotted as gas pressure $P_{\rm gas} = \rho c_{\rm s}^2$) for the $30$ most-collapsed cores in each of the simulations. Similar to Fig.~\ref{fig:sim_profile}. The most collapsed core in each simulation, which we analyzed in this work, is plotted in blue.}
\label{fig:allprofiles}
\end{figure*}

We also look at how the first-collapsed cores in each simulation compare to a larger sample of turbulent fragmentation cores. 
Fig.~\ref{fig:allprofiles} shows the radial density profiles of the $30$ most-collapsed cores in each of the simulations.
We see that there is variation among the normalization of the core profiles (as expected, since the turbulent fragmentation process is known to yield a powerlaw distribution of masses). The first-collapsed cores are not outliers in terms of their total mass. From Fig.~\ref{fig:allprofiles} we see that some of the less-collapsed cores have a density profile shallower than $\rho\propto r^{-2}$ towards their centers; but reach $\rho\propto r^{-2}$ as they collapse further. This is true regardless of the mean-field strength.

\end{document}